\documentclass[aps,prd,10pt,showpacs,preprintnumbers,nofootinbib,superscriptaddress,notitlepage]{revtex4-1}

\usepackage{amsmath}
\usepackage{amssymb}
\usepackage{latexsym}
\usepackage{amsfonts}
\usepackage{epsfig}
\usepackage{psfrag}
\usepackage{graphicx}
\usepackage{wasysym}
\usepackage{ulem}
\usepackage{xcolor}
\usepackage{subfigure}
\usepackage{dsfont}

\usepackage{subfigure}

\usepackage{multirow}
\newcommand{\Eqref}[1]{Eq.~\eqref{#1}}

\renewcommand{\vec}[1]{\mathbf{#1}}

\begin{document}

\title{Photon merging and splitting in electromagnetic field inhomogeneities}

\author{Holger Gies}
\author{Felix Karbstein}
\author{Nico Seegert}
\affiliation{Theoretisch-Physikalisches Institut, Abbe Center of Photonics,
Friedrich-Schiller-Universit\"at Jena, Max-Wien-Platz 1, D-07743 Jena, Germany}
\affiliation{Helmholtz-Institut Jena, Fr\"obelstieg 3, D-07743 Jena, Germany}

\begin{abstract}
We investigate photon merging and splitting processes in
inhomogeneous, slowly varying electromagnetic fields. Our study is
based on the three-photon polarization tensor following from the
Heisenberg-Euler effective action.  We put special emphasis on
deviations from the well-known constant field results, also revisiting
the selection rules for these processes. In the context of
high-intensity laser facilities, we analytically determine compact
expressions for the number of merged/split photons as obtained in the
focal spots of intense laser beams. For the parameter range of a
typical petawatt class laser system as pump and a terawatt class laser
as probe, we provide estimates for the numbers of signal photons
attainable in an actual experiment. The combination of frequency
upshifting, polarization dependence and scattering off the
inhomogeneities renders photon merging an ideal signature for the
experimental exploration of nonlinear quantum vacuum properties.
\end{abstract}

\date{\today}

\pacs{12.20.Fv}

\maketitle

%%%%%%%%%%%%%%%%%%%%%%%%%%%
\section{Introduction}
%%%%%%%%%%%%%%%%%%%%%%%%%%% 

The vacuum of quantum electrodynamics (QED) acquires properties akin
to those of ordinary polarizable matter when subjected to strong
electromagnetic fields
\cite{Euler:1935zz,Heisenberg:1935qt,Weisskopf}. These fields can
couple to electron-positron fluctuations inducing nonlinear
interactions (for reviews, see
\cite{Dittrich:1985yb,Dittrich:2000zu,Marklund:2008gj,Dunne:2008kc,Heinzl:2008an,%
  DiPiazza:2011tq,Dunne:2012vv,Battesti:2012hf,King:2015tba}). The
spatial and temporal scale associated to these electron-positron fluctuations
is set by the Compton wavelength $\lambda_C = 1/m \approx
3.86 \times 10^{-13}{\rm m}$ and the Compton time $\tau_C = 1/m
\approx 1.29 \times 10^{-21}{\rm s}$ of the electron respectively,
where $m \approx 511{\rm keV}$ denotes the electron mass. At leading
order in the field strengths prominent signatures of quantum vacuum
nonlinearities such as vacuum magnetic birefringence
\cite{Toll:1952,Baier,BialynickaBirula:1970vy} and direct
light-by-light scattering \cite{Euler:1935zz,Karplus:1950zz} are
mediated by an effective four-photon interaction.

The resulting nonlinear interactions among laboratory electromagnetic
fields are suppressed by powers of the field strength ratio
$\mathcal{E}/\mathcal{E}_{\rm cr}$, with $\cal E$ denoting the
electric/magnetic field amplitude and $\mathcal{E}_{\rm cr} = m^2/e
\approx 1.3 \cdot 10^{18}{\rm V/m} \approx 4\cdot 10^9{\rm T}$ the
critical field strength. Hence, experimental verifications
\cite{Akhmadaliev:1998zz,Akhmadaliev:2001ik} have so far been limited
to high-energy experiments probing vacuum nonlinearities in the strong
Coulomb fields in the vicinity of highly charged ions
(Delbr\"uck-scattering \cite{Delbrueck:1933,Bethe:1952yya} and
  photon splitting \cite{Adler:1971wn,Lee:1998hu}). Vacuum
nonlinearities in macroscopic electromagnetic fields have not been
directly verified so far. Direct searches of vacuum magnetic
birefringence in macroscopic magnetic fields
\cite{Cantatore:2008zz,Berceau:2011zz} for instance, have already
demonstrated that a combination of both high field strength as well
as a high signal detection sensitivity will eventually be needed for
a first discovery, see \cite{Zavattini:2016sqz} for a recent
proposal.

On the other hand, recent technological advances in the development of
high-intensity lasers have begun to access new extreme-field
territory. The perspective to directly probe quantum vacuum
nonlinearities in all-optical pump-probe type set-ups is most
promising: one high-intensity laser (``pump'') generates a strong
electromagnetic field pulse polarizing the quantum vacuum in its
focus, which is then probed by a second high-intensity laser. A
prominent example is the fundamental physics program within the HIBEF
project \cite{HIBEF}, which plans to combine a near-infrared petawatt
(PW) laser as pump and the European XFEL as probe laser aiming to
detect vacuum birefringence
\cite{Heinzl:2006xc,Karbstein:2015xra,Schlenvoigt:2016}, see also
\cite{DiPiazza:2006pr,Dinu:2013gaa} for related work. While the
advantage of using ultra-intense lasers is obvious from the
accessible field strengths, important progress also has been made on
the detection side in the form of high-purity x-ray polarimetry
\cite{Uschmann:2014}.  Further attractive theoretical proposals
have focused on optical signatures of quantum vacuum nonlinearities
based on interference effects
\cite{King:2013am,Tommasini:2010fb,Hatsagortsyan:2011}, photon-photon
scattering in the form of laser-pulse collisions
\cite{Lundstrom:2005za,Lundin:2006wu,King:2012aw} and quantum
reflection \cite{Gies:2013yxa}.

In some of these studies, a new ingredient in addition to
extreme fields and high detection efficiencies has been identified:
as laser pulses feature an intrinsic spatio-temporal
structure, these spacetime inhomogeneities allow for a richer
variety of quantum vacuum signatures that remains invisible in the
idealized theoretical limit of constant homogeneous fields.
 
To this end, recently a representation of the one-loop photon polarization tensor
$\Pi^{\mu\nu}(k,k')$ (photon two-point function) in the limit of low
energies and momenta and for slowly varying but otherwise arbitrary
electromagnetic field inhomogeneities has been derived
\cite{Karbstein:2015cpa}. Note that essentially all macroscopic
electromagnetic fields attainable in the laboratory fall into this
category, as they vary on scales much larger than the Compton
wavelength and time of the electron. This result facilitates a
straightforward investigation of photon propagation effects in
high-intensity laser fields. Whereas constant electromagnetic fields
only affect a probe photon's polarization properties, inhomogeneous
fields may additionally alter its frequency and wave vector, which can
be employed to physically separate the (tiny) amount of photons
carrying the signature from a large background consisting of probe
photons unaffected by quantum vacuum nonlinearities;
inhomogeneities thus become the key ingredient for some phenomena
such as quantum reflection \cite{Gies:2013yxa}.

Besides propagation effects,
photons in strong electromagnetic fields can also experience splitting
\cite{BialynickaBirula:1970vy,Adler:1971wn,Adler:1970gg,Papanyan:1971cv,Papanyan:1973xa,Stoneham:1979,Baier:1986cv,Baier:1996bq,Adler:1996cja,DiPiazza:2007yx}
and merging \cite{Yakovlev:1966,DiPiazza:2007cu,Gies:2014jia}. Photon
splitting describes processes where a single photon splits into two or
more outgoing photons under the influence of an external
electromagnetic field. Photon merging can be viewed as the inverse
process: two or more photons merge under the influence of the external
field, yielding a single outgoing photon. The first detailed
investigation of photon splitting has been performed by Adler in 1971
\cite{Adler:1971wn}, who considered this process in a constant, purely
magnetic background field.

In this work we aim at adopting the strategy devised in
\cite{Karbstein:2014fva,Karbstein:2015cpa,Karbstein:2015xra} to study
photon splitting and merging in the strong, inhomogeneous
electromagnetic fields attainable with high-intensity laser
experiments.  For this, we start in Sect.~\ref{sec:poltensor} by
deriving the ``three-photon polarization tensor''
$\Pi^{\mu\nu\rho}(k,k',k'')$ to one-loop order in the limit of low
energies and momenta and for weakly-varying but otherwise arbitrary
electromagnetic field inhomogeneities. This quantity describes the
effective interaction between three photon fields facilitated by
vacuum fluctuations in the presence of external electromagnetic
fields. It accounts for couplings to the external field to all orders.
We furthermore detail on how to determine the amplitudes and numbers
of signal photons for the splitting and merging processes. In
Sect.~\ref{sec:polcrossed}, we analyze the polarization properties and
selection rules for photon splitting and merging. More specifically,
we focus on the special class of field inhomogeneities characterized
by unidirectional orthogonal electric and magnetic fields of equal
strength, but featuring arbitrary field amplitude
profiles. In Sect.~\ref{sec:mergsplit}, we consider a specific field
amplitude profile and compare the magnitudes of the numbers of photons
experiencing photon splitting and merging in this field inhomogeneity.
Finally, we provide estimates for the number of accessible signal
photons from the photon merging process for realistic laser
parameters.  We finish with a conclusion in Sect.~\ref{sec:concl}.

%%%%%%%%%%%%%%%%%%%%%%%%%%%
\section{The three-photon polarization tensor in an electromagnetic field inhomogeneity}		\label{sec:poltensor}
%%%%%%%%%%%%%%%%%%%%%%%%%%% 

\begin{figure}[htpb]
\centering
 \includegraphics[width=0.7\textwidth]{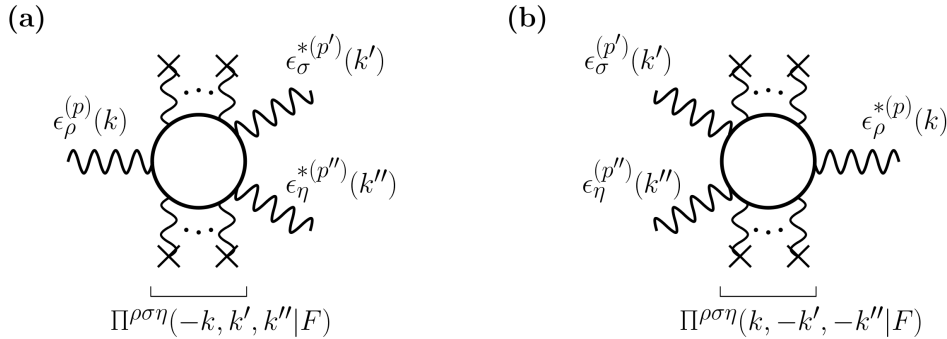}
\caption{\small Feynman diagrams of the three-photon interactions considered in this work. (a) Photon Splitting: An incoming photon with momentum $k^\mu$ and polarization four-vector $\epsilon_{\rho}^{(p)}(k)$ splits, under the assistance of the external field inhomogeneity $F^{\mu\nu}(x)$, into two photons with $\bigl\{{k'}^\mu$, $\epsilon_{\sigma}^{*(p')}(k')\bigr\}$ and $\bigl\{{k''}^\mu$, $\epsilon_{\eta}^{*(p'')}(k'')\bigr\}$. (b) Photon Merging: Two incoming photons with $\bigl\{{k'}^\mu$, $\epsilon_{\sigma}^{(p')}(k')\bigl\}$ and $\bigl\{{k''}^\mu$, $\epsilon_{\eta}^{(p'')}(k'')\bigl\}$ merge to yield a single outgoing photon with $\bigl\{k^\mu$, $\epsilon_{\rho}^{*(p)}(k)\bigl\}$. The coupling of the probe photons and the external field inhomogeneity $F^{\mu\nu}(x)$ to the electron-positron loop is encoded in the three-photon polarization tensor $\Pi^{\rho\sigma\eta}$.}
\label{fig:SplittingMergingFeyncon}
\end{figure}

The goal of this article is to study photon splitting and merging in
an all-optical experiment, where both pump and probe fields are
provided by lasers. We consider a macroscopic ``probe'' photon field
$a_{\rho}(q)$ traversing an electromagnetic ``pump`` field
configuration described by the field-strength profile $F^{\mu\nu}(x)$.
The nonlinear interactions between the pump and probe fields may
induce outgoing signal photons via the effective interactions of the
probe and pump fields. Each signal photon is characterized by its
four-momentum $k^\mu$ and polarization four-vector
$\epsilon_\sigma^{*(p)}(k)$, where $p$ labels the two transverse
photon polarizations. For photon splitting we study
the process linear in the probe photon field giving rise to two
outgoing signal photons.  Conversely, for photon merging we consider
the process quadratic in the probe photon field resulting in a single
signal photon. More specifically, the amplitude $\mathcal{M}_{\rm
  Split}^{p \rightarrow p'p''}(k',k'')$ for photons from a macroscopic
probe photon field $a_\rho^{(p)}(q)$, with momentum $q^\mu$ and
polarization vector $\epsilon_\rho^{(p)}(q)$, to split into two real
photons with momenta ${k'}^{\mu}=(|\vec{k}'|,\vec{k}')$ and
${k''}^\mu=(|\vec{k}''|,\vec{k}'')$, and polarization vectors
$\epsilon_\sigma^{*(p')}(k')$ and $\epsilon_\eta^{*(p'')}(k'')$
respectively, is given by
\begin{equation}
 \mathcal{M}_{\rm Split}^{p \rightarrow p'p''}(k',k'') = \frac{ \epsilon_\sigma^{*(p')}(k') }{\sqrt{2|\vec{k}'|}} \, \frac{ \epsilon_\eta^{*(p'')}(k'') }{\sqrt{2|\vec{k}''|}} \int_{q} \Pi^{\rho\sigma\eta}(-q,k',k''|F)a_{\rho}^{(p)}(q).	\label{eq:AmplSplit}
\end{equation}
Here, $\Pi^{\rho\sigma\eta}(-q,k',k''|F)$ denotes the three-photon
polarization tensor (three-point proper vertex) in an external
electromagnetic field inhomogeneity $F^{\mu\nu}(x)$, and the indices
$p,p',p''$ label the polarizations of the incident probe photon beam
and the signal photons to be specified later. We use ``all-outgoing''
sign conventions for the momentum arguments of $\Pi^{\rho\sigma\eta}$.
At one loop-order, the three-photon polarization tensor quantifies the
effective coupling of three photon fields mediated by an
electron-positron loop. At this stage, $\Pi^{\rho\sigma\eta}$ accounts
for the coupling to the field inhomogeneity $F^{\mu\nu}(x)$ to all
orders, see Fig. \ref{fig:SplittingMergingFeyncon}.  Throughout the
paper we work in Heaviside-Lorentz units, setting $\hbar=c=1$.
Spatio-temporal four vectors are denoted by italic letters, $x^{\mu} =
(t,{\rm x},{\rm y},{\rm z})$, for the spatial components we use roman
letters.  Our metric is $g^{\mu\nu}=(-,+,+,+)$, such that $kx \equiv
k_{\mu}x^{\mu} = (-k^0 t + \vec{k} \cdot \vec{x})$. We employ the
short-hand notation $\int_q \equiv \int \frac{{\rm d}^4q}{(2\pi)^4}$
for momentum integrations, and $\int_x \equiv \int {\rm d}^4x$ for
space-time integrations.

In analogy to Eq. (\ref{eq:AmplSplit}), the amplitude
$\mathcal{M}^{p'p''\rightarrow p}_{\rm Merg}(k)$ for merging photons
from the probe field to yield a single outgoing photon with momentum
$k^\mu = (|\vec{k}|,\vec{k})$ and polarization vector
$\epsilon_\rho^{*(p)}(k)$ reads
\begin{equation}
 \mathcal{M}^{p'p''\rightarrow p}_{\rm Merg}(k) = \frac{ \epsilon_\rho^{*(p)}(k) }{\sqrt{2|\vec{k}|}} \, \int_{q'} \int_{q''} \Pi^{\rho\sigma\eta}(k,-q',-q''|F) a'_{\sigma}{}^{(p')}(q') a''_{\eta}{}^{(p'')}(q'').	\label{eq:AmplMerg}
\end{equation}
Here we accounted for the most generic situation, where the merged
photons are originating from two distinct probe photon fields
$a'_{\sigma}{}^{(p')}(q')$ and $a''_{\eta}{}^{(p'')}(q'')$ with
polarizations $p'$ and $p''$, respectively.

The physical information about the splitting and merging processes in
Eqs.~\eqref{eq:AmplSplit} and \eqref{eq:AmplMerg} is encoded in the
three-photon polarization tensor.  At present, no exact analytical
results for the three-photon polarization tensor for arbitrary
momentum transfers and background field inhomogeneities are known.  At
one-loop order and for generic momentum transfers, exact results for
two classes of background field configurations are available.  The
first class comprises uniform electromagnetic fields: Papanyan and
Ritus \cite{Papanyan:1971cv,Papanyan:1973xa} derived a parameter
integral representation of the three-photon polarization tensor in
``constant-crossed'' fields, i.e. for orthogonal constant electric and
magnetic fields of equal amplitude $|\vec{E}|=|\vec{B}|=:\mathcal{E}$.
For this setting, the field invariants $\mathcal{F} = \frac{1}{4}
F_{\mu\nu}F^{\mu\nu} = \frac{1}{2} (\vec{B}^2-\vec{E}^2)$ and
$\mathcal{G} = \frac{1}{4} F_{\mu\nu} {}^*F^{\mu\nu} =
-\vec{E}\cdot\vec{B}$ vanish; ${}^*F^{\mu\nu} =
\frac{1}{2}\epsilon^{\mu\nu\alpha\beta}F_{\alpha\beta}$ denotes the
dual field strength tensor.  Later, splitting amplitudes valid for
arbitrary photon energies have been derived in constant, but
generically oriented electromagnetic fields \cite{Baier:1986cv}.
Energy-momentum conservation in constant fields implies that all three
photons propagate collinearly. As a consequence, an expansion of the
amplitude in the background field strength starts with terms
$\propto\left(\frac{e\mathcal{E}}{m^2}\right)^3$, i.e., arising from a
{\it hexagon diagram} in accordance with the Adler theorem
\cite{Adler:1971wn}. This will also become manifest in our results
below.

The second class of background field configurations encompasses
plane-wave backgrounds with vector potential $A_{\mu}(\kappa x)$,
where $\kappa^\mu = (|\pmb{\kappa}|,\pmb{\kappa})$ denotes the
four-momentum of the plane wave.  Photon splitting amplitudes for
arbitrary photon energies in this class of background fields have been
derived in \cite{DiPiazza:2007yx}.  Recall that photon splitting
amplitudes amount to on-shell matrix elements of the three-photon
photon polarization tensor.  In this configuration, the probe photons
can exchange energy and momentum with the background field, and an
expansion in the background field strength hence generically starts
with terms $\propto\left(\frac{e\mathcal{E}}{m^2}\right)$,
corresponding to \textit{box diagrams}. For small angles between the
photon momenta, splitting amplitudes have also been calculated in
Coulomb fields \cite{Lee:1998hu}.

A study of photon splitting and merging in generic background field
inhomogeneities may eventually require dedicated numerical efforts,
see, e.g., \cite{worldline}. However, in the present work we confine
ourselves to field inhomogeneities whose spatial and temporal
variation, $w$ and $\tau$, are large compared to the spatial and
temporal Compton scale of the electron-positron loop, i.e.,
$w\gg\lambda_C$ and $\tau\gg\tau_C$. These \textit{slowly varying
  fields} mark the regime of the locally constant field approximation
(LCFA). As a matter of fact, practically all current and proposed
all-optical probes of QED vacuum nonlinearities in macroscopic fields
fall into this category.

The LCFA corresponds to treating the microscopic quantum fluctuations
as propagating in a constant field at each point in spacetime. This
culminates in the effective Heisenberg-Euler Lagrangian $\mathcal{L}(x|F)$ which is a
local function of spacetime. On the
macroscopic level, this Lagrangian can now be applied locally. In the
present work, we use it to calculate the three-photon polarization
tensor for manifestly inhomogeneous background field profiles
following the strategy of \cite{Karbstein:2015cpa}. For strictly
constant background fields, an analogous idea has originally been
adopted by Adler \cite{Adler:1971wn} and Bialynicka-Birula
\cite{BialynickaBirula:1970vy} in the 1970s to study photon
splitting. Still, it is important to realize that the same line or
reasoning also applies to slowly varying fields with parametrically
suppressed errors which are quantified below. The photon polarization
tensor derived in this way is manifestly restricted to low
four-momentum transfers; i.e., also the probe photons have to be
slowly varying. Let us emphasize that this procedure manifestly
preserves gauge invariance and therefore retains complete information
about the polarization properties of the involved photons for any given
slowly varying background field inhomogeneity.

Our starting point is the one-loop effective action $\mathcal{S}_{\rm
  int}(F,f)=\int_x \mathcal{L}(x|F+f)$, mediating
effective interactions between probe photon fields $a_\mu(x)$, with
field strength tensor $f^{\mu\nu}(x)$, in a background field
inhomogeneity with field strength tensor $F^{\mu\nu}(x)$. It is
obtained from the Heisenberg-Euler Lagrangian for constant fields,
which only depends on the field strength tensor
$F^{\mu\nu}$,\footnote{In fact, the Heisenberg-Euler Lagrangian
  depends on $F^{\mu\nu}$ only via the field invariants $\cal F$ and
  $\cal G$.}  by substituting $F^{\mu\nu}\to
F^{\mu\nu}(x)+f^{\mu\nu}(x)$,
cf.~\cite{Karbstein:2015cpa}. Correspondingly, the effective
interactions among the probe photons can be seen as generated by
derivatives of the Heisenberg-Euler Lagrangian.  An expansion in terms
of probe photons reads
\begin{equation}
 \mathcal{S}_{\rm int} = \mathcal{S}^{(1)}_{\rm int} + \mathcal{S}^{(2)}_{\rm int} + \mathcal{S}^{(3)}_{\rm int} + ... 
	    = \int_x \frac{\partial \mathcal{L}}{\partial F^{\mu\nu} } f^{\mu\nu}  \, + \frac{1}{2!} \int_x \frac{\partial^2 \mathcal{L}}{\partial F^{\mu\nu} \partial F^{\alpha\beta} } f^{\mu\nu} f^{\alpha\beta} \, + \frac{1}{3!} \int_x \frac{\partial^3 \mathcal{L}}{\partial F^{\mu\nu} \partial F^{\alpha\beta} \partial F^{\gamma\delta} } f^{\mu\nu} f^{\alpha\beta} f^{\gamma\delta} \, + ... \quad ,	\label{eq:ActionExpandend}
\end{equation}
where $S_{\rm int}^{(l)}$ contains the effective interaction between
$l$ photons.  To keep the notation compact, we have omitted the
spacetime arguments of the fields. As a consequence of Furry's
theorem, only an even number of contractions of $F^{\mu\nu}(x)$ and
$f^{\mu\nu}(x)$ may constitute building blocks of the Lagrangian. On
the level of the action, the LCFA neglects derivatives of the field
strength tensors which are generic constituents of the exact effective
Lagrangian.  Derivatives of the fields in position space translate to
multiplications of the fields in momentum space with their typical
momentum $v$. As the action is dimensionless, the LCFA therefore
inherently neglects contributions of $\mathcal{O}\bigl(
\frac{v^2}{m^2} \bigr)$, with the electron mass $m$ as the only
dimensionful scale in QED.

The lowest order interaction term $\mathcal{S}^{(1)}_{\rm int}$ of the
expansion (\ref{eq:ActionExpandend}) describes vacuum emission
processes \cite{Karbstein:2014fva}. The second order term
$\mathcal{S}^{(2)}_{\rm int}$ entails propagation effects
\cite{Karbstein:2015cpa}, such as vacuum birefringence
\cite{Karbstein:2015xra} and quantum reflection
\cite{Gies:2013yxa}. Finally, the third order term
$\mathcal{S}^{(3)}_{\rm int}$ encodes three-photon interactions such
as splitting and merging to be considered here. The three-photon
polarization tensor in momentum space is now easily inferred from
$\mathcal{S}^{(3)}_{\rm int}$. Employing the momentum space
representation of the probe photons, $a^{\mu}(x) = \int_k {\rm
  e}^{ikx} a^{\mu}(k)$, we substitute $f^{\mu\nu}(x)=i \int_k {\rm
  e}^{ikx} \bigl[ k^{\mu} g^{\nu\sigma} - k^{\nu} g^{\mu\sigma}\bigr]
a_{\sigma}(k)$ into $\mathcal{S}_{\rm int}^{(3)}$ and obtain
\begin{equation}
 \mathcal{S}_{\rm int}^{(3)} = -\frac{1}{3!} \int_k \int_{k'} \int_{k''}  a_{\rho}(k) a_{\sigma}(k') a_{\eta}(k'') \Pi^{\rho\sigma\eta}(k,k',k''),
\end{equation}
where the three-photon polarization tensor has been defined as
\begin{multline}
 \Pi^{\rho\sigma\eta}(k,k',k'') := i (k^{\mu}g^{\nu\rho} - k^{\nu}g^{\mu\rho} ) (k'{}^{\alpha}g^{\beta\sigma} - k'{}^{\beta}g^{\alpha\sigma} )
 (k''{}^{\gamma}g^{\delta\eta} - k''{}^{\delta}g^{\gamma\eta} ) \int_x {\rm e}^{i(k+k'+k'')x} \frac{\partial^3 \mathcal{L}}{\partial F^{\mu\nu} \partial F^{\alpha\beta} \partial F^{\gamma\delta} }(x) .
\end{multline}
The tensorial structure guarantees that the Ward-identity, ensuring
gauge invariance, is fulfilled: $k_{\rho}
\Pi^{\rho\sigma\eta}(k,k',k'') = k'_{\sigma}
\Pi^{\rho\sigma\eta}(k,k',k'') = k''_{\eta}
\Pi^{\rho\sigma\eta}(k,k',k'') = 0$. This three-photon polarization
tensor neglects contributions of order $v^3
\mathcal{O}\bigl(\frac{v^2}{m^2}\bigr)$.

For explicit calculations it is useful to rewrite the derivatives of
the Lagrangian with respect to $F^{\mu\nu}$ in terms of the invariants
$\mathcal{F}$ and $\mathcal{G}$.  Then, the third derivative of the
Lagrangian is spanned by 20 independent tensor structures,
corresponding to the full set of basis elements of a completely
symmetric rank-3 spacetime tensor. It reads
\begin{multline}
 \frac{\partial^3 \mathcal{L}}{\partial F^{\mu\nu} \partial F^{\alpha\beta} \partial F^{\gamma\delta} } =  \frac{1}{8} \biggl\{  \bigl[ \bigl( g_{\mu\alpha}g_{\nu\beta} - g_{\mu\beta}g_{\nu\alpha} \bigr) F_{\gamma\delta} + \bigl( g_{\mu\gamma}g_{\nu\delta} - g_{\mu\delta}g_{\nu\gamma} \bigr) F_{\alpha\beta} + \bigl( g_{\alpha\gamma}g_{\beta\delta} - g_{\gamma\beta}g_{\delta\alpha} \bigr) F_{\mu\nu} \bigr] \frac{\partial^2 \mathcal{L}}{\partial \mathcal{F}^2} \\
 + \bigl[ \epsilon_{\mu\nu\alpha\beta} {}^*F_{\gamma\delta} + \epsilon_{\mu\nu\gamma\delta} {}^*F_{\alpha\beta} + \epsilon_{\alpha\beta\gamma\delta} {}^*F_{\mu\nu} \bigr] \frac{\partial^2 \mathcal{L}}{\partial \mathcal{G}^2} \\
 + \bigl[ \bigl( g_{\mu\alpha}g_{\nu\beta} - g_{\mu\beta}g_{\nu\alpha} \bigr) {}^*F_{\gamma\delta} + \bigl( g_{\mu\gamma}g_{\nu\delta} - g_{\mu\delta}g_{\nu\gamma} \bigr) {}^*F_{\alpha\beta} + \bigl( g_{\alpha\gamma}g_{\beta\delta} - g_{\gamma\beta}g_{\delta\alpha} \bigr) {}^*F_{\mu\nu} \\
 + \epsilon_{\mu\nu\alpha\beta} F_{\gamma\delta} + \epsilon_{\mu\nu\gamma\delta} F_{\alpha\beta} + \epsilon_{\alpha\beta\gamma\delta} F_{\mu\nu} \bigr] \frac{\partial^2 \mathcal{L}}{\partial \mathcal{F} \partial \mathcal{G}}  \\
 + F_{\mu\nu} F_{\alpha\beta} F_{\gamma\delta}  \frac{\partial^3 \mathcal{L}}{\partial \mathcal{F}^3}  +  {}^*F_{\mu\nu} {}^*F_{\alpha\beta} {}^*F_{\gamma\delta}  \frac{\partial^3 \mathcal{L}}{\partial \mathcal{G}^3}  + \bigl[F_{\mu\nu} F_{\alpha\beta} {}^*F_{\gamma\delta} +  F_{\mu\nu} {}^*F_{\alpha\beta} F_{\gamma\delta} +  {}^*F_{\mu\nu} F_{\alpha\beta} F_{\gamma\delta} \bigr] \frac{\partial^3 \mathcal{L}}{\partial\mathcal{F}^2 \partial\mathcal{G}} \\ 
 + \bigl[{}^*F_{\mu\nu} {}^*F_{\alpha\beta} F_{\gamma\delta} +  F_{\mu\nu} {}^*F_{\alpha\beta} {}^*F_{\gamma\delta} +  {}^*F_{\mu\nu} F_{\alpha\beta} {}^*F_{\gamma\delta} \bigr] \frac{\partial^3 \mathcal{L}}{\partial\mathcal{F}  \partial\mathcal{G}^2} \biggr\}. \label{eq:d3LF^3}
\end{multline}

An explicit representation of the one-loop Heisenberg-Euler Lagrangian for constant electric and magnetic fields of arbitrary orientation and amplitudes is given in terms of a proper-time integral \cite{Heisenberg:1935qt,Schwinger:1951nm,Dittrich:2000zu},
\begin{equation}
 \mathcal{L}(\mathcal{F},\mathcal{G}) = \frac{\alpha}{2\pi} \int_0^{\infty} \frac{{\rm d}s}{s} \, {\rm e}^{-i \frac{m^2 s}{e}} \left[ |\mathcal{G}|\, \coth(as) \cot(bs) + \frac{2}{3}\mathcal{F} - \frac{1}{s^2} \right],	\label{eq:HE}
\end{equation}
where $a=\bigl( \sqrt{\mathcal{F}^2+\mathcal{G}^2}-\mathcal{F}
\bigr)^{\frac{1}{2}}$ and $b=\bigl(
\sqrt{\mathcal{F}^2+\mathcal{G}^2}+\mathcal{F} \bigr)^{\frac{1}{2}}$.
Due to CP invariance, this Lagrangian is an even function of $\cal G$.
The derivatives with respect to $\mathcal{F}$ and $\mathcal{G}$ in
\Eqref{eq:d3LF^3} can now be calculated from \Eqref{eq:HE}. For either
purely electric or purely magnetic fields, or alternatively for
orthogonal electric and magnetic fields the field invariant
$\mathcal{G}$ vanishes, and the proper-time integrals can be performed
analytically. This leads to a representation of the one-loop
Heisenberg-Euler Lagrangian and its derivatives in terms of $\Gamma$-
and Hurwitz $\zeta$-functions
(cf. \cite{Karbstein:2015cpa,Dittrich:1985yb,Dittrich:2000zu,Tsai:1975iz}). Hence,
for this class of configurations explicit analytical insights into the
strong-field limit are possible.

In the following, we concentrate on the case of crossed-fields with
$\vec{E}\cdot\vec{B} = 0$ and $|\vec{E}| = |\vec{B}| =:
\mathcal{E}$. This configuration is of particular importance as it can
be employed to describe the electromagnetic fields delivered by
high-intensity lasers; cf. Section \ref{sec:mergsplit}. Since both
field invariants vanish in this case, $\mathcal{F}=\mathcal{G}=0$, it
is useful to perform a weak field expansion of the Heisenberg-Euler
Lagrangian (\ref{eq:HE}),
\begin{equation}
 \mathcal{L}(\mathcal{F},\mathcal{G}) = \frac{\alpha}{90\pi} \left( \frac{e}{m^2} \right)^{2} \bigl( 7\mathcal{G}^2 + 4 \mathcal{F}^2 \bigr) - \frac{2 \alpha}{315\pi} \left( \frac{e}{m^2} \right)^{4} \bigl( 13 \mathcal{F} \mathcal{G}^2 + 8 \mathcal{F}^3 \bigr) + \mathcal{O}\left( \{\mathcal{F},\mathcal{G}\}^4 \right).	\label{eq:HEexpanded}
\end{equation}
The derivatives of the Lagrangian with respect to $\mathcal{F}$ and $\mathcal{G}$ are then given by
\begin{equation}
 \begin{split}
  & \frac{\partial \mathcal{L} }{\partial \mathcal{F} } = \frac{\partial \mathcal{L} }{\partial \mathcal{G} } = \mathcal{O} \bigl( \{\mathcal{F},\mathcal{G} \} \bigr), \\
  & \left\{ \frac{\partial^2 \mathcal{L} }{\partial \mathcal{F}^2 } , \frac{\partial^2 \mathcal{L} }{\partial \mathcal{G}^2 }, \frac{\partial^2 \mathcal{L} }{\partial \mathcal{F} \partial\mathcal{G} } \right\} = \bigl\{4,7,0 \bigr\} \frac{\alpha}{45\pi} \left( \frac{e}{m^2} \right)^{2} + \mathcal{O} \bigl( \{\mathcal{F},\mathcal{G} \} \bigr),\\
   & \left\{ \frac{\partial^3 \mathcal{L} }{\partial \mathcal{F}^3 } , \frac{\partial^3 \mathcal{L} }{\partial \mathcal{G}^3 }, \frac{\partial^3 \mathcal{L} }{\partial \mathcal{F} \partial\mathcal{G}^2} , \frac{\partial^3 \mathcal{L} }{\partial \mathcal{F}^2 \partial\mathcal{G} } \right\} = -\bigl\{24,0,13,0 \bigr\} \frac{4\alpha}{315\pi} \left( \frac{e}{m^2} \right)^{4} + \mathcal{O} \bigl( \{\mathcal{F},\mathcal{G} \} \bigr). \label{eq:deriv}
 \end{split} 
\end{equation}
For generic slowly varying backgrounds, neglecting higher-order
expansion terms is equivalent to a weak-field limit, $\frac{e
  \mathcal{E}}{m^2} \ll 1$, with $\mathcal{E}$ being a characteristic
field strength scale of the background. Working with
  Eqs.~\eqref{eq:HEexpanded}, \eqref{eq:deriv} corresponds to the same
  level of accuracy as has recently been used for a study of vacuum
  higher-harmonic generation in a slowly varying background
  \cite{King:2014vha} or constant crossed-field background in the
  shock regime \cite{Bohl:2015uba} based on the quantum equations of
  motion. For general crossed-field configurations considered 
here, higher-order terms vanish identically and the terms written
explicitly in \Eqref{eq:deriv} constitute the full result within the
LCFA. The corresponding parametric analysis can be made more
rigorously: for $\mathcal{F}=\mathcal{G}=0$, the different
contributions to the three-photon polarization tensor scale as $\sim m
\bigl(\frac{v}{m}\bigr)^3
\frac{e\mathcal{E}}{m^2}\bigl[1+\mathcal{O}\bigl(\frac{v^2}{m^2}\bigr)\bigr]$
for the term linear in $\cal E$, and as $\sim m \frac{v}{m}
\frac{e\mathcal{E}}{m^2} \bigl[ \bigl(\frac{e\mathcal{E}}{m^2}\bigr)^2
  \mathcal{O}\bigl(\frac{v^2}{m^2}\bigr)\bigr]^n$, with $n \in
\mathds{N}_+$, for higher powers of $\cal E$.  As the LCFA adopted
here neglects contributions
$\sim\mathcal{O}\bigl(\frac{v^2}{m^2}\bigr)$, terms with $n> 1$ are
not accounted for in the corresponding three-photon polarization
tensor. Hence, in the limit of $\mathcal{F}=\mathcal{G}=0$ we find
\begin{equation}
 \Pi^{\rho\sigma\eta}(k,k',k'') = i\frac{\alpha}{45\pi} \left( \frac{e}{m^2} \right) \int_x {\rm e}^{i(k+k'+k'')x} \biggl\{  \biggl( \frac{e\mathcal{E}{(x)}}{m^2} \biggr) c_{(1)}^{\rho\sigma\eta}(k,k',k'') - \frac{4}{7} \biggl( \frac{e\mathcal{E}{(x)}}{m^2} \biggr)^{3} c_{(3)}^{\rho\sigma\eta}(k,k',k'') \biggr\}.	\label{eq:3Polcrossed}
\end{equation}
Here, we have decomposed the polarization tensor into components
linear and cubic in the field amplitude $\cal E$,
\begin{eqnarray}
 c_{(1)}^{\rho\sigma\eta}(k,k',k'')  &=& 4 \Bigl[ \bigl( kk'g^{\rho\sigma}-k^{\sigma}k'{}^{\rho} \bigr) (k''\hat{F})^{\eta} +  \bigl( kk''g^{\rho\eta}-k^{\eta}k''{}^{\rho} \bigr) (k'\hat{F})^{\sigma} +  \bigl( k'k''g^{\sigma\eta}-k'{}^{\eta}k''{}^{\sigma} \bigr) (k\hat{F})^{\rho} \Bigr]  \nonumber\\
 &&- 7 \Bigl[ k_{\mu}k'_{\nu}\epsilon^{\mu\nu\rho\sigma} (k'' {}^*\hat{F})^{\eta} +  k_{\mu}k''_{\nu}\epsilon^{\mu\nu\rho\eta} (k' {}^*\hat{F})^{\sigma} + k'_{\mu}k''_{\nu}\epsilon^{\mu\nu\sigma\eta} (k {}^*\hat{F})^{\rho} \Bigr],	\label{eq:polfunct1} 
\\
 c_{(3)}^{\rho\sigma\eta}(k,k',k'')  &=& 24 (k\hat{F})^{\rho} (k'\hat{F})^{\sigma} (k''\hat{F})^{\eta} \nonumber\\ 
&& + 13 \Bigl[( k{}^*\hat{F})^{\rho} (k'{}^*\hat{F})^{\sigma} (k''\hat{F})^{\eta} +  (k\hat{F})^{\rho} (k'{}^*\hat{F})^{\sigma} (k''{}^*\hat{F})^{\eta} +  (k{}^*\hat{F})^{\rho} (k'\hat{F})^{\sigma} (k''{}^*\hat{F})^{\eta} \Bigr], 	\label{eq:polfunct3}
\end{eqnarray}
employing the short-hand notation $(k\hat{F})^{\rho} :=
k_\nu\hat{F}^{\nu\rho}$.  Note that \Eqref{eq:3Polcrossed} is spanned
by just 10 tensor structures, as further 10 tensor structures have
been eliminated by the 10 independent equations of the Ward identity.
Additionally, we have introduced the
normalized field strength tensor $\hat{F}^{\mu\nu}$ as $F^{\mu\nu} =
\hat{F}^{\mu\nu} \mathcal{E}$, which is independent of $x$ for \textit{unidirectional} fields, i.e.,
$\vec{E}=\hat{\vec{e}}_{\rm E}\,{\cal E}(x)$ and $\vec{B}=\hat{\vec{e}}_{\rm
  B}\,{\cal E}(x)$.  This is, e.g., the case for linearly polarized
Gaussian laser beams in the paraxial approximation.

For inhomogeneities with sufficiently simple profiles, the space-time
integration can be performed straightforwardly. For constant
backgrounds $\mathcal{E}(x)=\mathcal{E}$, this yields delta functions
$(2\pi)^4\delta^{(4)}(k+k'+k'')$ in \Eqref{eq:3Polcrossed} which
enforce energy and momentum conservation.  It is particularly
instructive to study the constant-field limit for the case where
$k^\mu$, $k'^\mu$ and $k''^\mu$ are four-momenta describing real
photons. In this case, we have $k^\mu=\hat{k}^\mu \omega$, with photon
frequency $\omega=|\vec{k}|$ and normalized four-momentum
$\hat{k}^\mu=(1,\hat{\vec{k}})$. The unit vector
$\hat{\vec{k}}=\vec{k}/\omega$ points into the photon propagation
direction; in turn $\hat{k}^\mu$ is frequency independent, solely
representing the propagation geometry. Accounting for the relative
sign for in- and outgoing photons [cf. Eqs.~\eqref{eq:AmplSplit} and
  \eqref{eq:AmplMerg}], all three photons propagate collinearly in the
constant-field case, and the combination $\delta^{(4)}(-k+k'+k'')
c_{(1)}^{\rho\sigma\eta}(-\hat{k},\hat{k}',\hat{k}'')=\delta^{(4)}(k-k'-k'')
c_{(1)}^{\rho\sigma\eta}(\hat{k},-\hat{k}',-\hat{k}'')$ vanishes.
Hence, in constant fields the lowest-order contributions to
three-photon amplitudes are of
$\mathcal{O}\bigl((\tfrac{e\mathcal{E}}{m^2})^{3}\bigr)$, which is a
manifestation of the Adler theorem \cite{Adler:1971wn}. Beyond the
constant-field limit, however, first-order contributions are expected
to become relevant if inhomogeneities facilitate an appreciable
four-momentum transfer between the background field and the probe
photons. This can give rise to interactions among probe photons whose
directions of propagation differ notably from each other.

In order to explicitly evaluate the number of signal photons from
splitting or merging from the amplitudes (\ref{eq:AmplSplit}) and
(\ref{eq:AmplMerg}), we have to specify the incoming photon fields.
The differential number ${\rm d}\mathcal{N}$ of induced photons from
photon splitting or merging can then be obtained from the
corresponding amplitude $\mathcal{M}$ by Fermi's Golden Rule, ${\rm
  d}^6\mathcal{N}^{p \rightarrow p'p''}_{\rm Split} = \frac{{\rm
    d}^3k'}{(2\pi)^3} \frac{{\rm d}^3k''}{(2\pi)^3} \bigl|
\mathcal{M}_{\rm Split}^{p \rightarrow p'p''}(k',k'') \bigr|^2$ and
${\rm d}^3\mathcal{N}^{p'p'' \rightarrow p}_{\rm Merg} = \frac{{\rm
    d}^3k}{(2\pi)^3} \bigl| \mathcal{M}_{\rm Merg}^{p'p'' \rightarrow
  p}(k) \bigr|^2$. In this work, we limit ourselves to incoming probe
photon beams modeled as monochromatic linearly polarized plane waves,
$a_\nu^{(p)}(x)=\frac{1}{2}\frac{\mathfrak
  E}{\omega}\epsilon_{\nu}^{(p)}(\hat{k})\,{\rm
  e}^{i\omega(\hat{k}x)}$. For plane waves, we can express the field
strength of the probe beams through the time averaged intensity,
$\mathfrak{E} = \sqrt{2\langle I \rangle}$, which in turn is related
to the photon current density $J = \frac{N}{\sigma T}$ (i.e. the
number of photons $N$ passing through an area $\sigma$ in a certain
time interval $T$) via $\langle I \rangle = \omega J$. Hence, the
formulae for the differential number of photons induced from either
photon splitting or merging can be compactly represented as
\begin{eqnarray}
 {\rm d}^6\mathcal{N}^{p \rightarrow p'p''}_{\rm Split} &=& J \frac{{\rm d}^3k'}{(2\pi)^3} \frac{{\rm d}^3k''}{(2\pi)^3} \left|\frac{\epsilon_\rho^{(p)}(k)}{\sqrt{2\omega}}   \Pi^{\rho\sigma\eta}(-k,k',k'') \frac{ \epsilon_\sigma^{*(p')}(k') }{\sqrt{2\omega'}} \, \frac{ \epsilon_\eta^{*(p'')}(k'') }{\sqrt{2\omega''}}  \right|^2 \, ,	\label{eq:dNSlpitplane}\\
 {\rm d}^3\mathcal{N}^{p'p'' \rightarrow p}_{\rm Merg} &=& \, J' \, J'' \, \frac{{\rm d}^3k}{(2\pi)^3} 
  \left| \frac{ \epsilon_\rho^{*(p)}(k) }{\sqrt{2\omega}} \, \Pi^{\rho\sigma\eta}(k,-k',-k'')  \frac{ \epsilon_\sigma^{(p')}(k') }{\sqrt{2\omega'}} \frac{\epsilon_\eta^{(p'')}(k'')}{\sqrt{2\omega''}} \right|^2 . \label{eq:dNMergplane}
\end{eqnarray}

%%%%%%%%%%%%%%%%%%%%%%%%%%%
\section{Polarization properties in the crossed-field case}	\label{sec:polcrossed}
%%%%%%%%%%%%%%%%%%%%%%%%%%% 

\begin{figure}[htpb]
\center
\begin{subfigure}
  \centering
 \includegraphics[width=0.6\textwidth]{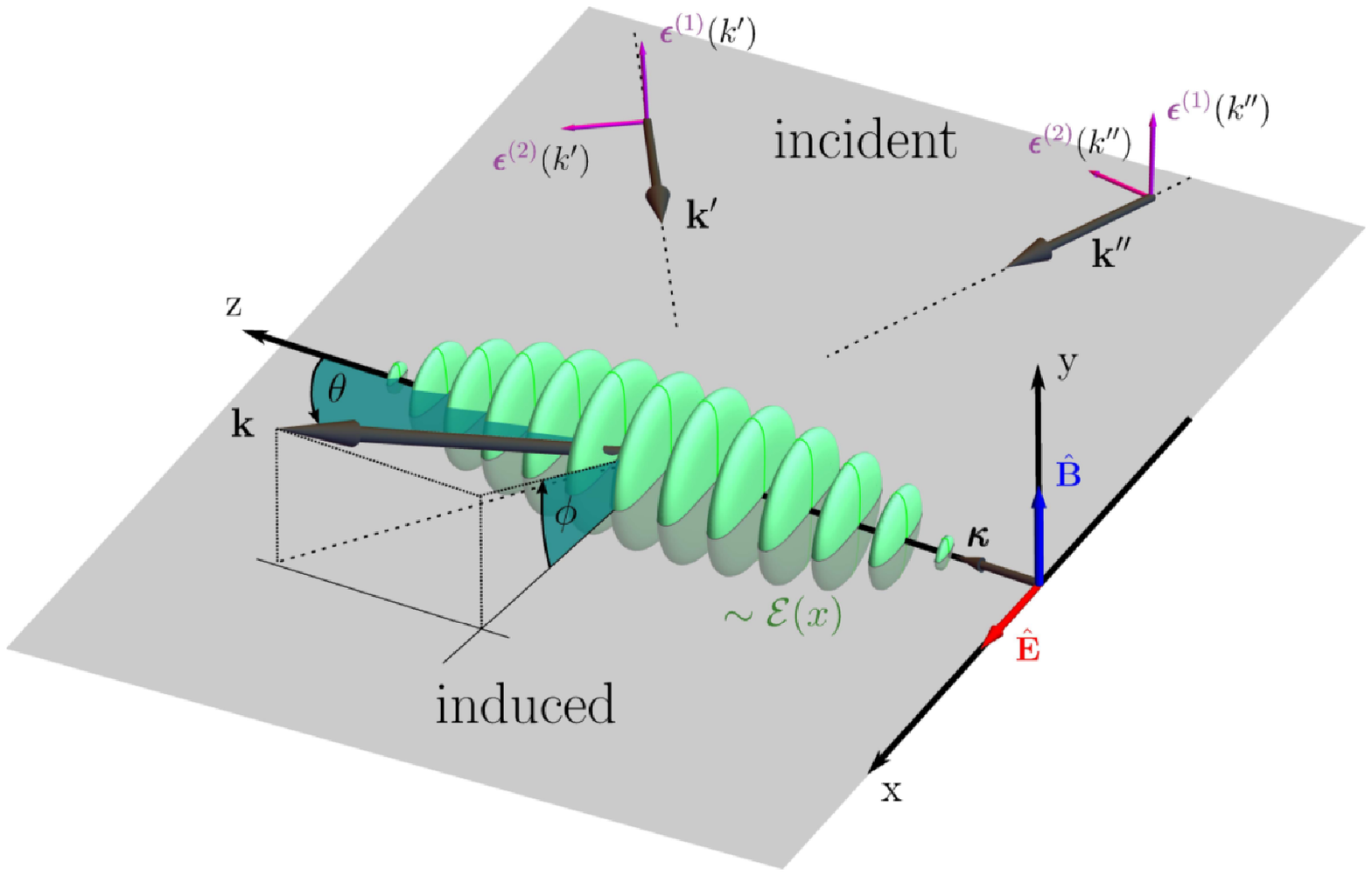}
\end{subfigure}
\hspace{0.15\textwidth}
\begin{subfigure}
  \centering
 \raisebox{0.2\height}{\includegraphics[width=0.15\textwidth]{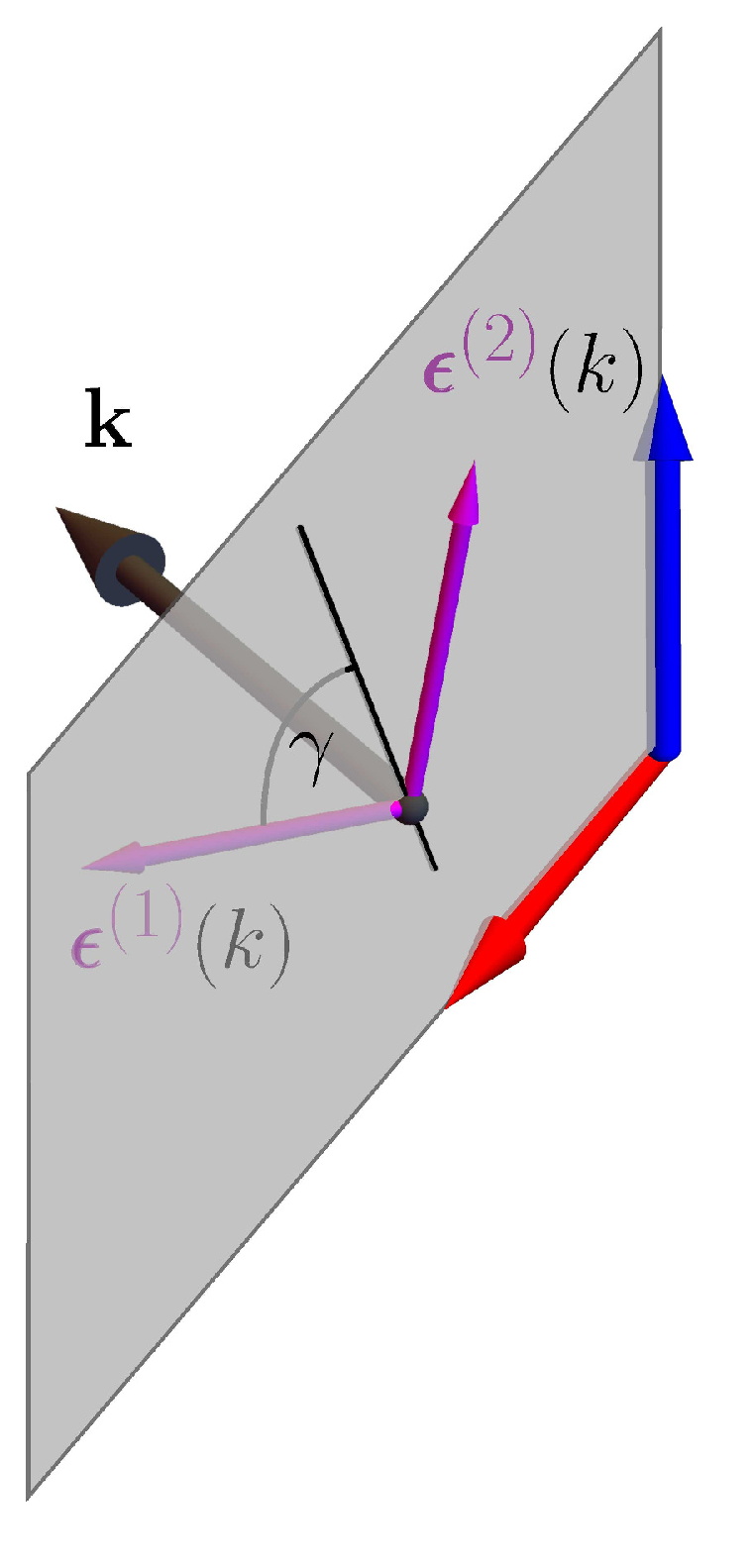}}
\end{subfigure}
\caption{\small Left: Geometry of photon merging in a localized
  background field inhomogeneity with mutually perpendicular
  $\vec{E}$, $\vec{B}$ and $\pmb{\kappa}\sim\hat{\vec{e}}_{\rm z}$ for
  the particular choice of $\varphi=0$ (cf. main text).  In this field
  configuration, two incoming probe photons with momenta $\vec{k}'$
  and $\vec{k}''$ may merge into one photon with momentum
  $\vec{k}$. For photon splitting (not depicted) the roles are
  reversed: An incident photon with momentum $\vec{k}$ may split into
  two photons with momenta $\vec{k}'$ and $\vec{k}''$. The
  polarization degrees of freedom of the photons are spanned by the
  unit vectors $\pmb{\epsilon}^{(1)}(q)$ and
  $\pmb{\epsilon}^{(2)}(q)$, where $q\in\{k,k',k''\}$. This figure
  depicts the special case where the incident photons propagate in the
  $\rm x$-$\rm z$ plane, and $\gamma'=\gamma''=0$. Right: The
  convention for the trihedron composed of $\hat{\vec{k}},\,
  \pmb{\epsilon}^{(1)}(k)$ and $\pmb{\epsilon}^{(2)}(k)$ is such that
  for $\gamma = 0$ the polarization vector $\pmb{\epsilon}^{(1)}(k)$
  lies in the $\rm x$-$\rm y$ plane (shaded area).}
\label{fig:Lasercon}
\end{figure}

For the remainder of this work we assume a unidirectional background
field inhomogeneity with orthogonal electric and magnetic fields,
characterized by the unit vectors $\hat{\vec{e}}_E$ and
$\hat{\vec{e}}_B$ respectively. We assume $\hat{\vec{e}}_E =
\cos\varphi \, \hat{\vec{e}}_{\rm x} + \sin\varphi \,
\hat{\vec{e}}_{\rm y}$ and $\hat{\vec{e}}_B = \hat{\vec{e}}_E
\bigr|_{\varphi \rightarrow \varphi + \frac{\pi}{2}}$. The direction of the field vectors in the ${\rm x}$-${\rm y}$ plane is
parametrized by the angle $\varphi \in [0,2\pi)$ (see
  Fig.~\ref{fig:Lasercon}). It is convenient to introduce the
  four-vector $\kappa^{\mu} = (1,\hat{\vec{e}}_E \times
  \hat{\vec{e}}_B) = (1,\hat{\vec{e}}_{\rm z})$; its spatial
  components correspond to the normalized Poynting vector of the
  background field. In order to describe the propagation and
  polarization properties of the probe photons efficiently, we switch to spherical
  coordinates. A probe photon's four-momentum is then given by $k^\mu =
  \omega (1, \hat{\vec{k}})$, where
  $\hat{\vec{k}}=(\cos\phi\sin\theta,\sin\phi\sin\theta,\cos\theta)$. Without
  loss of generality, we define its polarization four-vector for the
  polarization mode $p=1$ as $\epsilon_{\mu}^{(1)}(k) = \bigl(0,
  \pmb{\epsilon}^{(1)}(k) \bigr)$, with
\begin{equation}
  \pmb{\epsilon}^{(1)}(k) = \begin{pmatrix}
                       \cos\theta \, \cos\phi \, \sin\gamma - \sin\phi \, \cos\gamma	\\
                       \cos\theta \, \sin\phi \, \sin\gamma + \cos\phi \, \cos\gamma	\\
                       -\sin\theta \, \sin\gamma
                      \end{pmatrix}.
\end{equation}
It is straightforward to verify that $\hat{\vec{k}}$ and $
\pmb{\epsilon}^{(1)}(k)$ are orthogonal for any angle $\gamma$. The
corresponding second perpendicular polarization mode hence is given by
$\epsilon_{\mu}^{(2)}(k) = \epsilon_{\mu}^{(1)}(k) \bigr|_{\gamma
  \rightarrow \gamma - \frac{\pi}{2}}$. The photon state is now
completely characterized by its energy $\omega$ and the set of
parameters $\Theta := \left\{\theta,\phi,\gamma \right\}$. The angle
$\gamma \in [0,2\pi)$ determines the orientation of the trihedron
  composed of $\hat{\vec{k}},\, \pmb{\epsilon}^{(1)}(k)$ and
  $\pmb{\epsilon}^{(2)}(k)$. For $\gamma = 0$, the polarization vector
  $\pmb{\epsilon}^{(1)}(k)$ lies in the $\rm x$-$\rm y$ plane, while
  $\pmb{\epsilon}^{(2)}(k)$ lies in the plane spanned by
  $\hat{\vec{e}}_{\rm z}$ and $\hat{\vec{k}}$,
  cf. Fig.~\ref{fig:Lasercon}.  For $\theta=\{0,\pi\}$ both
  polarization vectors lie in the $\rm x$-$\rm y$ plane. Since we have
  left the angle $\gamma$ in the definition of
  $\pmb{\epsilon}^{(1)}(k)$ unspecified, it actually suffices to
  perform all the subsequent calculations exclusively for the choice
  of $p=p'=p''=1$ in order to study the interactions of linearly
  polarized photon beams.  All other linear photon polarizations can be
  addressed by shifting the angles $\gamma$, $\gamma'$ and
  $\gamma''$ accordingly.

Equations (\ref{eq:dNSlpitplane}) and (\ref{eq:dNMergplane}) require
us to calculate contractions of photon polarization vectors with the
polarization tensor. To this end, we define the {\it polarization
  overlap functions} as the contraction of the photon polarization
vectors with the tensor structures in Eqs.~\eqref{eq:polfunct1} and
\eqref{eq:polfunct3},
\begin{equation}
c^{pp'p''}_{(n)}(\Theta,\Theta',\Theta'',\varphi) := \epsilon_{\rho}^{(p)}(k) \epsilon_{\sigma}^{(p')}(k{}') \epsilon_{\eta}^{(p'')}(k{}'')   c_{(n)}^{\rho\sigma\eta}(\hat{k},\hat{k}',\hat{k}'').
\end{equation}
Recall, that the label $n$ refers to the contributions linear ($n=1$)
and cubic ($n=3$) in external field amplitude. We obtain
\begin{eqnarray}
& c^{111}_{(1)}(\Theta,\Theta',\Theta'',\varphi) =& 2 \sin^2\frac{\theta''}{2} \Bigl( \bigl[(1-\cos\theta\cos\theta')\cos(\phi-\phi') - \sin\theta\sin\theta' \bigr]\bigl[ 4 \sin\delta'' \cos(\gamma+\gamma') + 7 \cos\delta'' \sin(\gamma+\gamma') \bigr] \nonumber\\
&&\qquad \qquad + (\cos\theta-\cos\theta')\sin(\phi-\phi') \bigl[ 4\sin\delta'' \sin(\gamma+\gamma') - 7 \cos\delta'' \cos(\gamma+\gamma') \bigr] \Bigr) \nonumber\\
&& +  \text{cyclic perm. of } \Theta, \Theta', \Theta'' \ ,	\label{eq:poloverlap1}\\
& c^{111}_{(3)}(\Theta,\Theta',\Theta'',\varphi) =& -8 \sin^2\frac{\theta}{2} \sin^2\frac{\theta'}{2} \sin^2\frac{\theta''}{2} \Bigl(  24 \sin\delta \sin\delta' \sin\delta''  \nonumber\\ 
&&\qquad\qquad\qquad\qquad\qquad\quad + 13 \bigl[  \sin\delta \cos\delta' \cos\delta'' + \cos\delta \sin\delta' \cos\delta'' + \cos\delta \cos\delta' \sin\delta'' \bigr] \Bigr),	\label{eq:poloverlap2}
\end{eqnarray}
where $\delta := \varphi -\gamma-\phi$, and likewise for the primed quantities.

The polarization overlap functions are independent of the background
amplitude profile ${\cal E}(x)$.  Furthermore, they are fully
symmetric with respect to an exchange of the photons $\Theta$,
$\Theta'$ and $\Theta''$. If two photons, say $\Theta$ and $\Theta'$,
propagate parallelly, the term proportional to
$\sin^2\frac{\theta''}{2}$ in $c^{pp'p''}_{(1)}$ vanishes, as then
$\theta=\theta'$ and $\phi=\phi'$.  Consequently, if all three photons
propagate in the same direction, which is the case, e.g., for photon
splitting in constant fields, $c^{pp'p''}_{(1)}$ vanishes and the
three-photon amplitudes are cubic in the background field (Adler
theorem).  Also note that the polarization overlap functions
$c_{(1)}^{pp'p''}$ and $c_{(3)}^{pp'p''}$ behave quite differently
with regard to probe photon propagation along the direction of the
background field's normalized Poynting vector
${\pmb\kappa}=\hat{\vec{e}}_{\rm z}$.  In general, $c_{(1)}^{pp'p''}$
does not vanish if at least one photon's propagation direction differs
from $\hat{\vec{e}}_{\rm z}$. In contrast, $c_{(3)}^{pp'p''}$ vanishes
if at least one photon travels along $\hat{\vec{e}}_{\rm z}$.  In the
context of pure photon propagation effects, i.e., on the level of
single-photon to single-photon transition amplitudes in external
fields, it is a well-known fact that photon propagation is not
modified in weak crossed-field backgrounds, if
$(\hat{\vec{k}},\hat{\vec{e}}_E,\hat{\vec{e}}_B)$  form a basis of a
right-handed orthogonal coordinate system
\cite{Dittrich:2000zu}.  Since in our case the function
$c_{(3)}^{pp'p''}$ determines the polarization properties in constant-crossed
background fields to leading order in the background field, we find a
similar behavior for photon splitting and merging here.

Let us briefly investigate the selection rules for the photon merging
and splitting processes. These are a direct consequence of the
structure of the Heisenberg-Euler Lagrangian being an even function of
$\cal G$ as dictated by the CP invariance of QED.  This exerts a
strong influence on the structure of the three-photon polarization
tensor~\eqref{eq:3Polcrossed}; cf. also \Eqref{eq:d3LF^3}.

Considering the case of unidirectional backgrounds, the normalized
field strength tensor $\hat F^{\mu\nu}$ is $x$ independent.  For a
given four-momentum $k^\mu$, one can then construct four independent
four-vectors: $k^\mu$, $(\hat F^2k)^{\mu}$, $(k{}^*\hat{F})^{\mu}$ and
$(k\hat{F})^{\mu}$.  The latter two span the physical polarization
eigenmodes of a probe photon of momentum $k^\mu$.  We call the
  polarization $a^\mu \sim (k{}^*\hat{F})^{\mu}$ the ``slow'' (s) mode
  of propagation, since the external field reduces its phase
  velocity to $v_{\rm ph}^{(\text s)} \simeq 1-\frac{14}{45}
  \frac{\alpha}{4\pi}\rho (\frac{\mathcal{E}}{\mathcal{E}_{\rm
      cr}})^2$ (where $\rho$ is a purely geometrical factor
  \cite{Dittrich:2000zu}). Correspondingly, $(k\hat{F})^{\mu}$
  describes the ``fast'' (f) mode with $v_{\rm ph}^{(\text f)} \simeq
  1- \frac{8}{45} \frac{\alpha}{4\pi}\rho
  (\frac{\mathcal{E}}{\mathcal{E}_{\rm cr}})^2\geq v_{\rm ph}^{(\text
    s)}$. As
$(k{}^*\hat{F})^{\mu}(k\hat{F})_\mu=0$, the 10 tensor structures
(\ref{eq:polfunct1}) and (\ref{eq:polfunct3}) give rise to selection
rules which determine the allowed interactions between in- and
outgoing probe photons in the slow and fast mode.

\begin{table}[htpb]
 \centering
 \begin{tabular}{|p{4cm}||c|c||c||}
 \hline \multirow{2}{*}{(Splitting), (Merging)}					&\multicolumn{2}{c||}{Allowed?}	&	In $\rm x$-$\rm z$ plane, $\varphi=0$ 		\\
 \cline{2-4} 									&	$n=1$	&	$n=3$	&	$n=1,3$		\\
 \hline \hline (s $\rightarrow$ s$'$,s$''$) , (s$'$,s$''$ $\rightarrow$ s)	&	Yes	&	No	&	No	 	\\
 \hline (f $\rightarrow$ f$'$,f$''$) , (f$'$,f$''$ $\rightarrow$ f)		&	Yes	&	Yes	&	Yes		\\
 \hline (s $\rightarrow$ f$'$,f$''$) , (f$'$,f$''$ $\rightarrow$ s)		&		&		&			\\
    (f $\rightarrow$ s$'$,f$''$) , (s$'$,f$''$ $\rightarrow$ f)			&	Yes	&	No	&	No		\\
    (f $\rightarrow$ f$'$,s$''$) , (f$'$,s$''$ $\rightarrow$ f)			&		&		&			\\
 \hline (s $\rightarrow$ s$'$,f$''$) , (s$'$,f$''$ $\rightarrow$ s)		&		&		&			\\
    (f $\rightarrow$ s$'$,s$''$) , (s$'$,s$''$ $\rightarrow$ f)			&	Yes	&	Yes	&	Yes		\\
    (s $\rightarrow$ f$'$,s$''$) , (f$'$,s$''$ $\rightarrow$ s)			&		&		&			\\
    \hline
 \end{tabular}
  \caption{\small Selection rules for photon splitting and merging for generic 
propagation directions (middle column), as well as the special
case where all photons propagate in
the $\rm x$-$\rm z$ plane and the choice $\varphi=0$ for the
background field polarization (right column).  These selection rules
can be inferred from the tensor structures
$c_{(n)}^{\rho\sigma\eta}(k,k',k'')$ in Eqs.~\eqref{eq:polfunct1} and
\eqref{eq:polfunct3}. Here, ``s'' (``f'') denotes probe photons
polarized in the ``slow'' (``fast'') polarization mode in the
background field, see main text.  Photon splitting and merging are
governed by the same selection rules. The selection rules for
processes which are cubic in the background field strength ($n=3$)
agree with the well-known rules valid in the constant-field field
limit \cite{Papanyan:1973xa}. By contrast, first-order processes
($n=1$) generically lift the restrictions for the $n=3$ case, unless
the wave-vectors of all probe photons are confined to the $\rm x$-$\rm
z$ plane, i.e., for $\phi=\phi'=\phi''=\{0,\pi\}$, and $\varphi=0$
(right column). For this special case we have
$\epsilon_\mu^{(1)}(k)|_{\phi=\{0,\pi\},\gamma=0} \sim
(k{}^*\hat{F})_{\mu}$ and
$\epsilon_\mu^{(1)}(k)|_{\phi=\{0,\pi\},\gamma=\frac{\pi}{2}} \sim
(k\hat{F})_{\mu}$, such that the ``s'' (``f'') polarization mode
corresponds to the choice of $\gamma=0$ ($\gamma=\frac{\pi}{2}$),
cf. the main text. }
 \label{tab:selectionrules}
\end{table}

Table \ref{tab:selectionrules} lists the resulting selection
rules. Photon splitting and merging are both governed by the same
selection rules, since they are both inferred from the same tensor
structures $c_{(n)}^{\rho\sigma\eta}(k,k',k'')$,
cf.~Eqs.~(\ref{eq:polfunct1}) and (\ref{eq:polfunct3}). Processes
which are cubic in the background field ($n=3$) feature selection
rules which are well-known from photon splitting and merging in
constant fields \cite{Papanyan:1973xa}: only processes involving
either three fast, or one fast and two slow photons are permitted in
this case. 

However, inhomogeneous backgrounds allow for momentum transfers to
probe photons. This gives rise to processes linear in the background
field which potentially dominate
over processes cubic in the background. The function
$c_{(1)}^{\rho\sigma\eta}(k,k',k'')$ then determines the leading order
polarization properties, and the restrictions on the selection rules
of the process cubic in the background field are lifted, see Table
\ref{tab:selectionrules} (middle column).

For the explicit calculations performed in this work, we have employed
the four-vectors $\epsilon_\mu^{(p)}(k)$ to specify the polarization
states of the incoming and outgoing probe photons.  In contrast to
$(k{}^*\hat{F})^{\mu}$ and $(k\hat{F})^{\mu}$, the
$\epsilon_\mu^{(p)}(k)$ form a polarization basis independently of the
background.  Of course, we can always tune a given polarization vector
$\epsilon_\mu^{(1)}(k)$ to either the slow or the fast mode by
adjusting the angle $\gamma$.  Generically, the appropriate choice of
$\gamma$ depends on the propagation direction $\hat{\bf k}$ of the
considered photon. A notable exception is obtained by restricting the
photon propagation to the ${\rm x}$-${\rm z}$ plane and specializing
the background field polarization to $\varphi = 0$. In this case, the
background electric field points along $\hat{\vec{e}}_{\rm x}$, and
the magnetic field along $\hat{\vec{e}}_{\rm y}$ as in
Fig.~\ref{fig:Lasercon}. 
In this case we find that the choice of $\gamma=0$ 
coincides with the slow photon mode, as
$\epsilon_\mu^{(1)}(k)\big|_{\phi=\{0,\pi\},\gamma=0} \sim
(k{}^*\hat{F})_{\mu}$. Likewise, $\gamma=\frac{\pi}{2}$ describes the
fast photon mode, as
$\epsilon_\mu^{(1)}(k)\big|_{\phi=\{0,\pi\},\gamma=\frac{\pi}{2}} \sim
(k\hat{F})_{\mu}$. For this special case, we observe that the
selection rules for the processes linear and cubic in the background
field coincide; cf. Table \ref{tab:selectionrules} (right column).

%%%%%%%%%%%%%%%%%%%%%%%%%%%
\section{Photon merging and splitting in a localized background inhomogeneity}	\label{sec:mergsplit}
%%%%%%%%%%%%%%%%%%%%%%%%%%% 

In what follows, we specialize the background inhomogeneity to
resemble the electromagnetic field configuration in the focal spot of
a pulsed high-intensity laser beam, propagating along the $\rm z$
direction with normalized four-wavevector $\kappa^\mu$
(cf. Sec.~\ref{sec:polcrossed} above).  We assume a linearly polarized
beam with unidirectional perpendicular electric and magnetic fields of
equal amplitude profile (see Fig. \ref{fig:Lasercon}),
\begin{equation}
 \left( \frac{e\mathcal{E}(x)}{m^2}\right) = \left( \frac{e\mathcal{E}}{m^2}\right) {\rm e}^{-\bigl(\frac{2r}{w_0}\bigr)^2} {\rm e}^{-\bigl(\frac{2\rm z}{w_{\rm z}}\bigr)^2} {\rm e}^{-\bigl(\frac{2({\rm z}-t)}{\tau}\bigr)^2} \cos\bigl(\Omega({\rm z}-t)\bigr),	\label{eq:exponentialprofile}
\end{equation}
where $r=\sqrt{{\rm x}^2+{\rm y}^2}$.  Equation
(\ref{eq:exponentialprofile}) mimics the profile of a pulsed laser
beam of peak field amplitude $\cal E$, frequency $\Omega$ and pulse
duration $\tau$, which is focussed around $\rm z =0$.  The transversal
profile is a Gaussian with $\frac{1}{e}$-width $w_0$, resembling the
transversal profile of a Gaussian laser beam.  We neglect beam
divergence effects and assume this width to be constant along the
beam. This can be justified by the fact that the considered phenomena
become sizable only within the Rayleigh range of the focussed laser
beam. Here, beam widening effects are small, and the beam width can be
considered as approximately constant.  The length $w_{\rm z}/2$ mimics
the Rayleigh length of the pump laser beam. Note that the real
longitudinal profile of a Gaussian beam is the square root of a
Lorentzian ($\propto 1/\sqrt{1+(2{\rm z} /w_{\rm z})^2}$) rather than
a Gaussian, as has been chosen here. Nevertheless, the qualitative
features of photon splitting and merging are expected to result in
quantitatively comparable effects for both types of profiles. The
exponential profile simply helps us to obtain more compact formulae due to
the appearance of Gaussian integrals in Eq. (\ref{eq:3Polcrossed}).

Inserting the field profile Eq. (\ref{eq:exponentialprofile}) into the
polarization tensor Eq. (\ref{eq:3Polcrossed}), and plugging the
resulting expression into Eqs. (\ref{eq:dNSlpitplane}) and
(\ref{eq:dNMergplane}), yields the induced numbers of signal photons
due to photon splitting and merging,
\begin{eqnarray}
 \left\{ \begin{matrix}
          {\rm d}^6\mathcal{N}^{p \rightarrow p'p''}_{\rm Split}	\\
          {\rm d}^3\mathcal{N}^{p'p'' \rightarrow p}_{\rm Merg}
         \end{matrix}	\right\}
&=& 
  \left\{ \begin{matrix}
	  \frac{J}{(2\pi)^3}{\rm d}^3k'{\rm d}^3k''	\\
	  J' J''  {\rm d}^3k
         \end{matrix}	\right\}
\frac{w_0^4 w_{\rm z}^2 \tau^2 \alpha^2}{11520^2 \pi} \left( \frac{e\mathcal{E}}{m^2}\right)^2 \left( \frac{e }{m^2}\right)^2  \omega \omega' \omega'' \label{eq:MergedSplit}\\
&&\times \Biggl| c^{pp'p''}_{(1)}(\Theta,\Theta',\Theta'',\varphi) \sum_{\ell = \pm 1} {\rm e}^{-\frac{1}{16}\left[w_0^2 (q_{\rm x}^2 + q_{\rm y}^2) + w_{\rm z}^2(q^0-q_{\rm z})^2 + \tau^2(q^0+\ell \Omega)^2\right]} 
 \nonumber\\
&&\quad\,\, -\frac{1}{63} \left( \frac{e\mathcal{E}}{m^2}\right)^2  c^{pp'p''}_{(3)}(\Theta,\Theta',\Theta'',\varphi) \sum_{\ell = 0}^3 [1+\ell(3-\ell)] {\rm e}^{-\frac{1}{48}\left[w_0^2 (q_{\rm x}^2 + q_{\rm y}^2) + w_{\rm z}^2(q^0-q_{\rm z})^2 + \tau^2(q^0+(3-2\ell) \Omega)^2\right]} \Biggr|^2.	\nonumber
\end{eqnarray}
Here, $q^{\mu}:=k^\mu-k'^\mu-k''^\mu$ denotes the four-momentum
exchange of the probe and signal photons. The value of $q^{\mu}$
measures the deviation from the four-momentum conservation law in
constant background fields, $q^\mu|_{\text{const. bg.}}=0$.  The
contribution proportional to $c_{(1)}$ in Eq. (\ref{eq:MergedSplit})
encodes the process linear in the background: the pump laser field
exchanges a single photon of frequency $\Omega$ with the probe photon
fields.  The contribution proportional to $c_{(3)}$ encodes the
process cubic in the pump field amplitude. Here the exchange of three
pump photons facilitates possible energy transfers of
$\{3\Omega,\Omega,-\Omega,-3\Omega\}$ between the pump laser pulse and
the probe photon beams.  On the level of the three-photon amplitude,
the latter process is generically suppressed by a factor of $\left(
\frac{e\mathcal{E}}{m^2}\right)^2$ compared with the linear process.
However, the exponential suppression of splitting and merging as a
function of the four-momentum transfer is smaller for the cubic than
for the linear process, as is visible from the prefactors
($\frac{1}{48}$) vs.~($\frac{1}{16}$). This allows for kinematical
situations where the cubic processes dominate over the linear ones.

Furthermore, the momentum dependences in the exponentials in
Eq.~(\ref{eq:MergedSplit}) give first insights into the emission
characteristics of photon splitting and merging.  A maximum of induced
signal photons occurs for those energy and angle parameters which lead
to a vanishing argument of one of the exponential functions in
Eq.~(\ref{eq:MergedSplit}).  This happens if the interacting photons
fulfill energy conservation $\omega-\omega'-\omega'' + \ell \Omega =
0$, with $\ell\in\{-3,-1,1,3\}$.  A specific set of energies then
results in a relation for the corresponding polar angles
$\{\theta,\theta',\theta''\}$: $\omega \cos\theta+ \ell \Omega =
\omega'\cos\theta' + \omega''\cos\theta''$ from $\rm z$-momentum
conservation. Finally, the propagation characteristics in the $\rm
x$-$\rm y$ plane transversal to the pump laser beam are determined by
the corresponding momentum conservation $k_{\rm x} - k'_{\rm x} -
k''_{\rm x} = 0$ and $k_{\rm y} - k'_{\rm y} - k''_{\rm y} = 0$.

On the one hand, the microscopic amplitudes for photon splitting and
merging coincide, as they are fully determined by the three-photon
polarization tensor. On the other hand, completely different scaling
behaviors occur for the number of signal photons evaluated from
Eqs.~\eqref{eq:dNSlpitplane} and \eqref{eq:dNMergplane}: photon
merging is quadratic in the macroscopic probe photon fields, whereas
photon splitting is linear in the probe photon field. To linear order
in the pump field, the ratio of $\mathcal{N}_{\rm Split}$ to
$\mathcal{N}_{\rm Merg}$ thus scales as $\frac{\mathcal{N}_{\rm
    Split}}{\mathcal{N}_{\rm
    Merg}}\sim\bigl(\frac{v}{m}\bigr)^4\bigl(\frac{e\mathcal{E}_{\rm
    in}}{m^2}\bigr)^{-2}$; cf. also \Eqref{eq:MergedSplit}.  Here,
$\mathcal{E}_{\rm in}$ denotes the field strength of each incoming
probe photon field, and $v$ is the typical momentum scale of the probe
photons. In the following, we show that set-ups with high-intensity
lasers in the optical regime ($\frac{v}{m} \ll 1$) strongly favor the
merging process because of a substantially different phase space for the
signal photons as well as the scaling with the incoming probe photon
currents.

To make contact with an experimental set-up, we assume the
inhomogeneous pump field to be generated by a high-intensity laser,
which is focussed down to the diffraction limit (attainable with a
focusing aperture with $f^{\#}=1$). In this case, the diameter of the
pump beam in its focus is given by twice its wavelength $\lambda_{\rm
  pump}=\frac{2\pi}{\Omega}$, such that $w_0=2\lambda_{\rm
  pump}$. Likewise, we identify $w_{\rm z}$ with twice the Rayleigh
length of a Gaussian beam, i.e. $w_{\rm z} = 2{\rm z}_R = 2\pi
\lambda_{\rm pump}$. The pulse duration is given by $\tau = \tau_{\rm
  pump}$.  Assuming that the effective focal area contains $86\%$ of
the laser energy $W$ ($\frac{1}{e^2}$-criterion), we estimate the peak
field-strength of the pump as
\begin{equation}
 {\cal E}^2= 2 \langle I\rangle \approx 2 \frac{0.86\,W}{\tau\,\sigma}\,,
\label{eq:EBpump}
\end{equation}
with focal area $\sigma\approx\pi\lambda^2$.  We employ
\Eqref{eq:EBpump} to determine the field strength of the pump laser
for given laser parameters. The analogous relation for the probe
beams is used to determine the photon current densities $J$,
introduced in
Eqs.~\eqref{eq:dNSlpitplane}-\eqref{eq:dNMergplane}. Given the probe
laser parameters such as pulse energy $W_{\rm probe}$, frequency
$\omega_{\rm probe} = \frac{2\pi}{\lambda_{\rm probe}}$ and pulse
duration $\tau_{\rm probe}$, we obtain $J = \frac{0.86}{2\pi^2}
\frac{W_{\rm probe}}{\tau_{\rm probe} \lambda_{\rm probe}}$.
Note that the plane-wave probe picture is only fully adequate for $\tau_{\rm probe}\geq\tau_{\rm pump}$.
Otherwise corrections because of the finite time overlap of the pump and probe laser pulses have to be taken into account.

For a first estimate, we adopt the parameters of state-of-the-art
high-intensity laser facilities, namely two identical, fully
synchronized petawatt-class laser systems of wavelength $\lambda_{\rm
  beam}=800{\rm nm}\approx4.06{\rm eV}^{-1}$, and pulse duration
$\tau_{\rm beam}=25{\rm fs}\approx38.0{\rm eV}^{-1}$.  Note that these
parameters match the parameters of the laser system to be installed at
ELI-NP \cite{ELI}. We assume an energy of $W_{\rm beam}=25{\rm
  J}\approx1.56\cdot10^{20}{\rm eV}$ for each beam, which corresponds
to a laser power of $1{\rm PW}$ per beam.  Note that this can be
considered as a rather conservative estimate as the ELI-NP lasers are
designed as $10{\rm PW}$ systems.  One of these lasers is assumed to
constitute the pump, and the second one is assumed to be frequency
doubled and split into two probe beams of equal power.  The energy
loss for frequency-doubling is estimated as $50\%$, while the pulse
duration is considered as unaffected by the frequency-doubling
process. For each of the two probe beams we thus have $\tau_{\rm
  probe}=\tau_{\rm beam}$, $\lambda_{\rm
  probe}=\frac{1}{2}\lambda_{\rm beam}$ and $W_{\rm
  probe}=\frac{1}{4}W_{\rm beam}$.  Of course, the parameters of the
pump are $\tau_{\rm pump}=\tau_{\rm beam}$, $\lambda_{\rm
  pump}=\lambda_{\rm beam}$, and $W_{\rm pump}=W_{\rm beam}$. The fact
that such a set-up greatly favors photon merging over splitting
becomes obvious from the ratio of the induced signal photons,
$\frac{\mathcal{N}_{\rm Split}}{\mathcal{N}_{\rm Merg}} \sim
\frac{v^4}{\mathcal{E}_{\rm in}^2} \sim \frac{\omega_{\rm
    probe}^4}{\mathcal{E}_{\rm probe}^2} \sim 10^{-16}$ (cf. above).

For the remainder, we therefore exclusively focus on photon merging.
We only retain contributions in Eq.~(\ref{eq:MergedSplit}) which are
of first order in the pump, and neglect third-order terms. The latter
are sub-leading for the considered kinematical settings dictated by
requirement that the argument of the exponential function in the
second line of \Eqref{eq:MergedSplit} should vanish.  The total number
of merged photons arises from Eq.~(\ref{eq:MergedSplit}) (second line)
by integrating over all possible energies of the merged photons,
$\omega = 0...\infty$. Energy conservation requires the merged photon
to have a final energy of $\approx 2\omega_{\rm probe} \pm
\Omega=(4\pm1)\omega_{\rm beam}$, corresponding to the
absorption/emission of one photon with energy $\Omega$ from/to the
pump laser field. For the specific beam configuration considered
below, the argument in the exponential function of
\Eqref{eq:MergedSplit} can only vanish for the emission
process. Hence, the induced signal photons will predominantly be
emitted with an energy of $\omega \approx 2\omega_{\rm probe} - \Omega
=3\omega_{\rm beam} = 4.6{\rm eV}$. The fact that the signal photon is
an odd harmonic of the probe can be used for efficient filtering and
detection techniques for optimizing the signal-to-noise ratio.

\begin{figure}[htpb]
\center
  \begin{subfigure}
    \centering
    \includegraphics[width=0.55\textwidth]{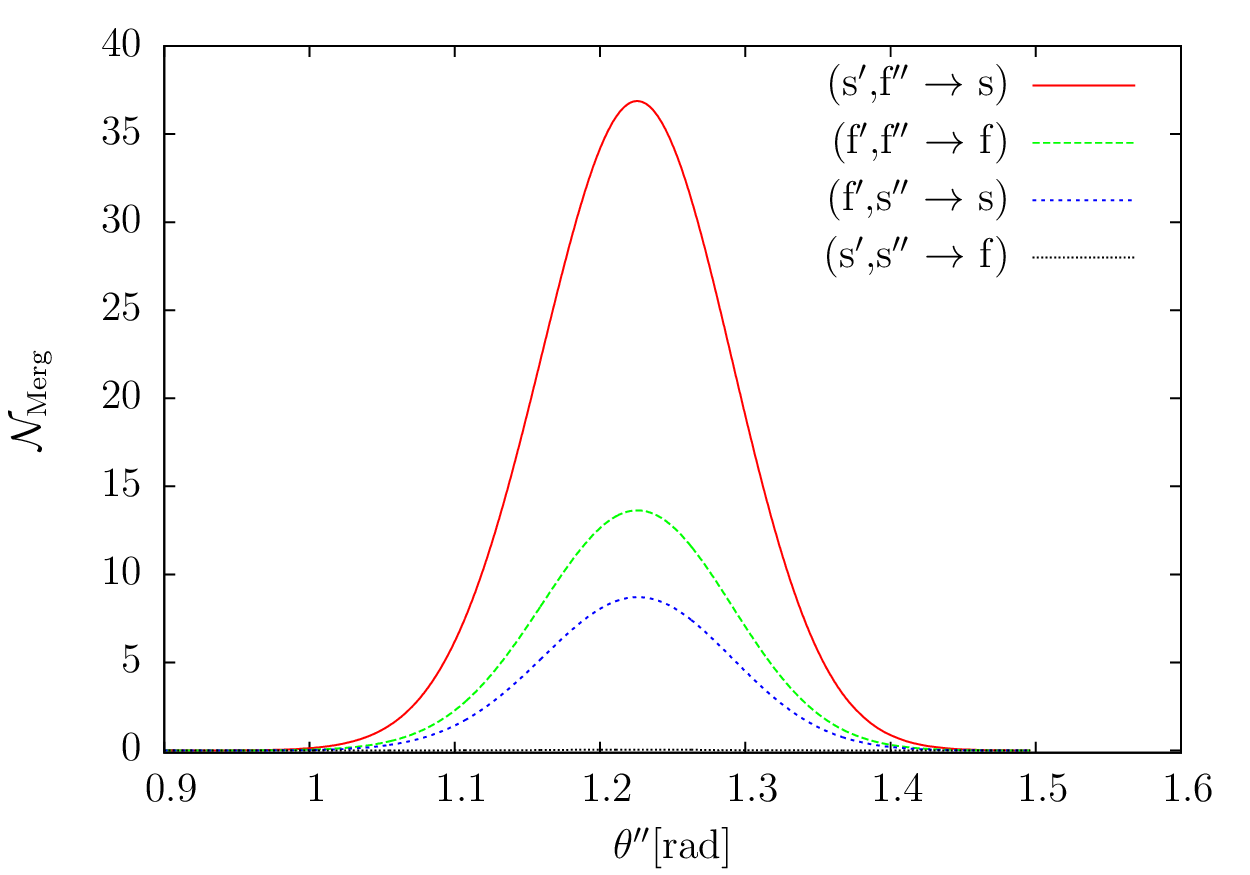}
  \end{subfigure}
  \begin{subfigure}
    \centering
    \raisebox{0.6\height}{\includegraphics[width=0.4\textwidth]{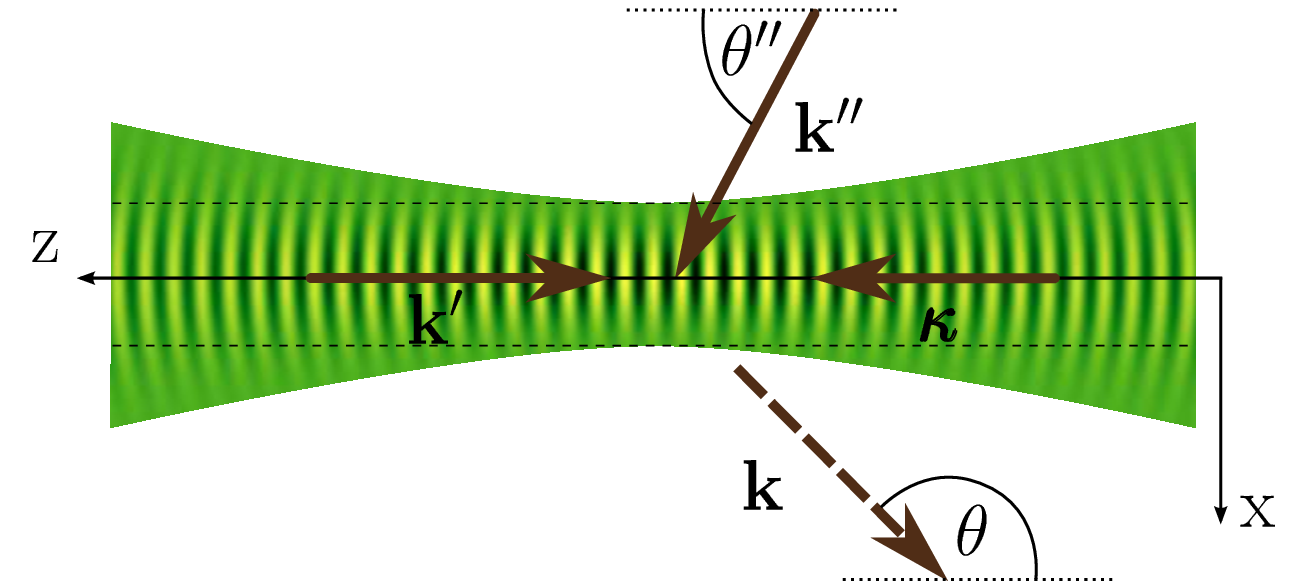}}
  \end{subfigure} 
\caption{\small Left panel: total number of merged photons
  $\mathcal{N}_{\rm Merg}$ as a function of the angle
  $\theta''=\sphericalangle(\vec{k}'',\hat{\vec{e}}_{\rm z})$
  attainable for the beam configuration sketched on right panel
  driven by two 1PW-class lasers as described in the main
  text. In the vicinity of the focal spot of a Gaussian beam,
  curvature effects can be neglected, justifying the simpler pump-beam
  profile (\ref{eq:exponentialprofile}) employed in this work
  (indicated by the dashed lines). The result includes an integration over the complete energy
  range and the full solid angle of the induced merged photons.  The
  probe beams have been chosen to propagate in the $\rm x$-$\rm z$
  plane, i.e. $\phi'=0$, $\phi''=0$, and $\theta' = \pi$ (cf. also
  Fig.~\ref{fig:Lasercon}). The polarization of the pump beam is
  chosen as $\varphi=0$. The plot depicts the number of merged photons
  for various polarization assignments of the probe photons, allowed
  according to the selection rules in Table \ref{tab:selectionrules}
  (right). For two 10PW driver lasers, the number of merged
  photons increases by a factor of 1000.}
\label{fig:FullMergingNumbercon}
\end{figure}

For simplicity, we limit ourselves now to pump and probe beam
propagation in the $\rm x$-$\rm z$ plane; cf. also
Tab.~\ref{tab:selectionrules} (right column).  Figure
\ref{fig:FullMergingNumbercon} shows the total number of merged
photons for the present set-up. Here, one probe beam with wave vector
$\vec{k}'$ counter-propagates the pump laser beam, and the second
probe beam with wave vector $\vec{k}''$ enters under an angle of
$\theta''$. The geometry is depicted in the right panel of
Fig.~\ref{fig:FullMergingNumbercon}. As is visible in the left panel
of Fig.~\ref{fig:FullMergingNumbercon}, this set-up yields a sizable
total number of merged photons near the optimum incoming angle of
$\theta'' \approx 1.23{\rm rad}$. The number of merged photons depends
strongly on the polarization modes of the probe photons, with a
maximum given for the parameter choice $\gamma=0$ for the merged
photon, and $\gamma'=0$ and $\gamma''=\frac{\pi}{2}$ for the probe
photons. As we limit ourselves to the $\rm x$-$\rm z$ plane and
$\varphi=0$, this choice can be identified with the process ${\rm s}',{\rm f}''\to{\rm s}$ (cf. Tab.~\ref{tab:selectionrules}).
As the number of merged signal photons scales as $\sim W_{\rm beam}^3$,
our results can straightforwardly be rescaled to the design
parameters of ELI-NP featuring two $10{\rm PW}$ laser beams by
multiplying with a factor of $1000$.

\begin{figure}[htpb]
\center
  \begin{subfigure}
    \centering
    \includegraphics[width=0.49\textwidth]{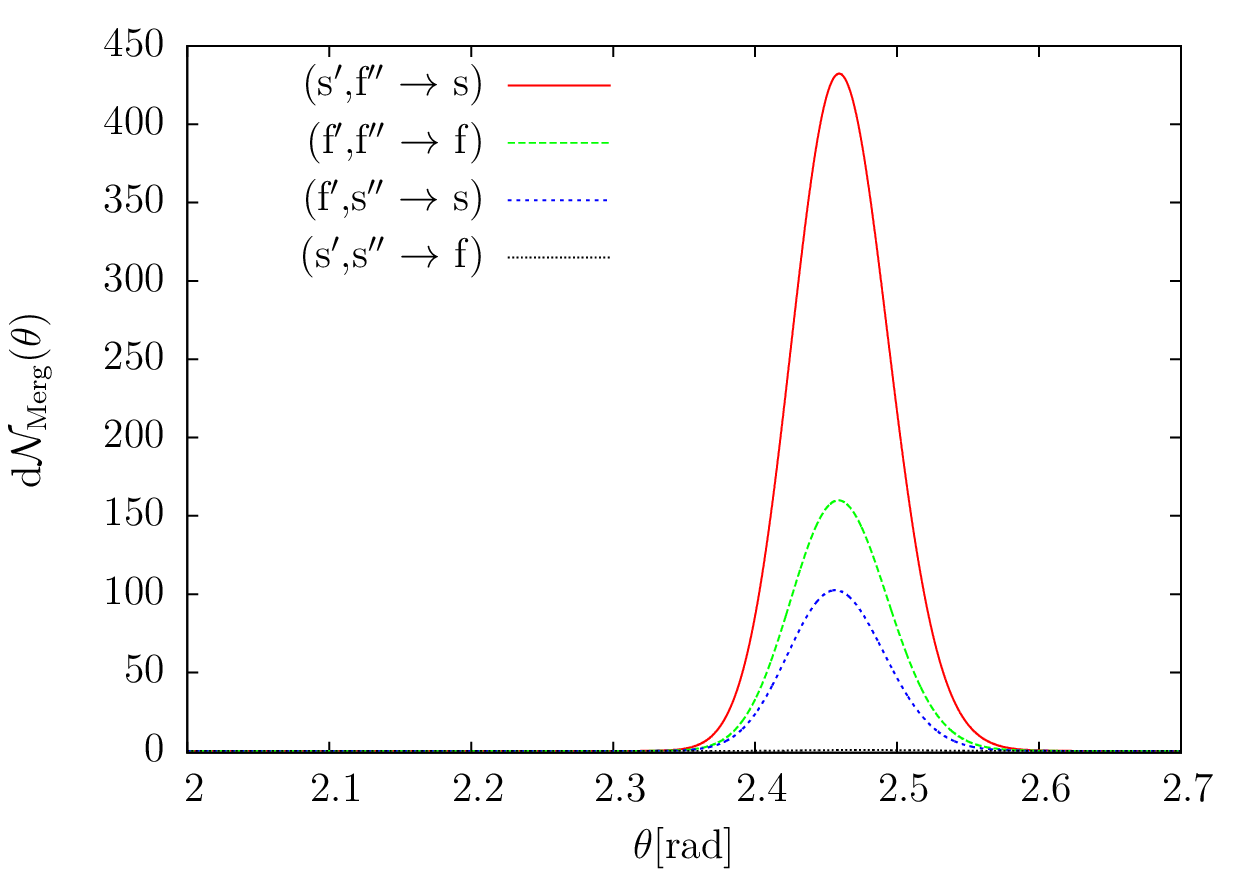}
  \end{subfigure}
  \begin{subfigure}
    \centering
    \includegraphics[width=0.49\textwidth]{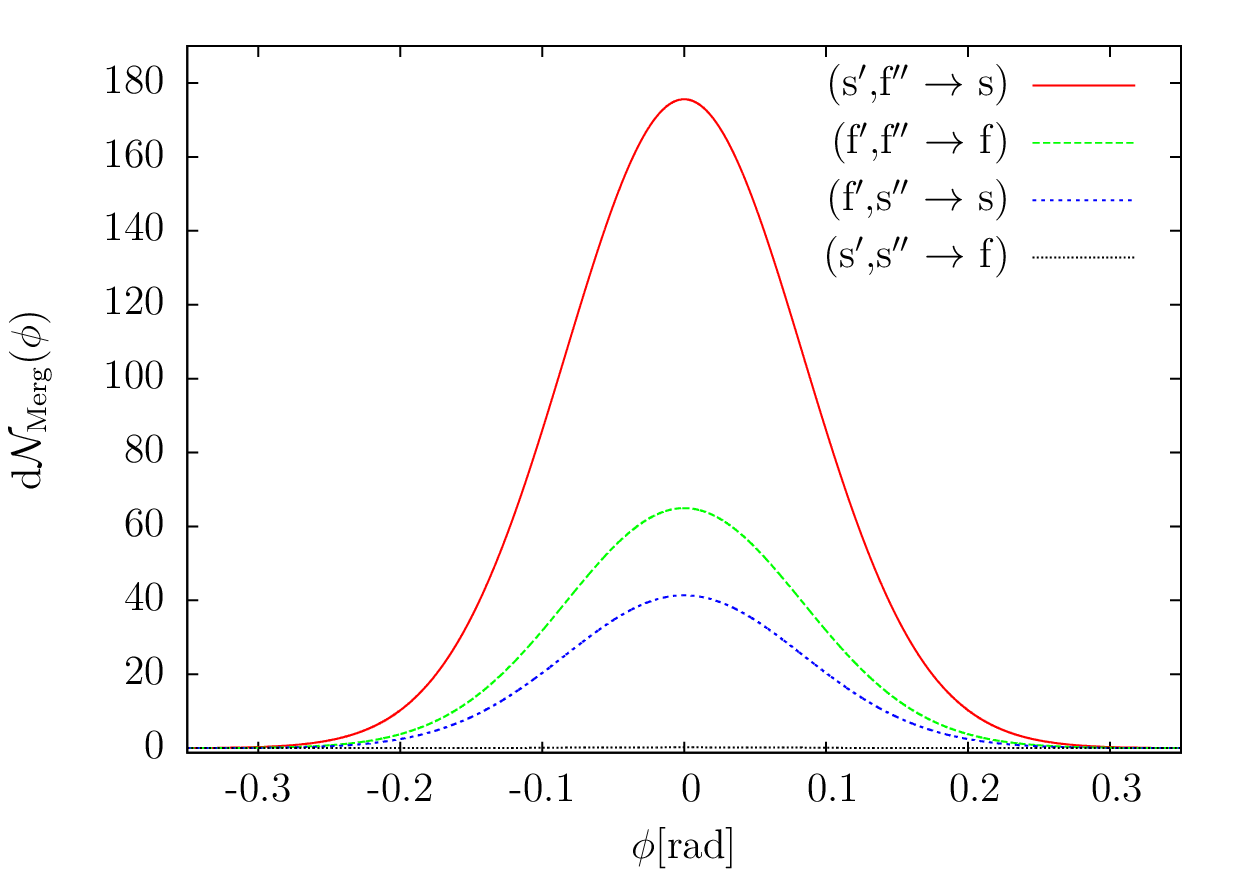}
  \end{subfigure}
  \caption{\small Emission characteristics of the attainable number of
    merged photons for our set-up driven by two 1PW-class
    lasers. The parameters of the incoming beams are chosen as
    $\theta'=\pi$, $\phi'=0$, $\theta''=1.23{\rm rad}$ and $\phi''=0$;
    the polarization of the pump beam is $\varphi=0$. The angle
    $\theta''$ maximizing the merged photon yield is visible in
    Fig.~\ref{fig:FullMergingNumbercon}.  Left panel: differential
    photon number ${\rm d}\mathcal{N}_{\rm Merg}(\theta)$ as a
    function of the polar angle $\theta$, with the corresponding
    energy and $\phi$ integrations performed over the full parameter
    regime.  Right panel: ${\rm d}\mathcal{N}_{\rm Merg}(\phi)$ as a
    function of the polar angle $\phi$, with the corresponding energy
    and polar angle integrals performed over their full parameter
    regimes. From these plots we infer that the merged signal photons
    are predominantly emitted in the $\rm x$-$\rm z$ plane, at an
    outgoing polar angle of $\theta \approx 2.46{\rm rad}$. For
    two 10PW driver lasers, the number of merged photons increases
    by a factor of 1000.}
\label{fig:OptMergingNumbercon}
\end{figure}

Figure \ref{fig:OptMergingNumbercon} displays the emission
characteristics for the optimum geometry with $\theta''=1.23{\rm rad}$
as inferred from Fig.~\ref{fig:FullMergingNumbercon}. The left panel
shows the distribution of merged photons as a function of the polar
angle $\theta$, while the right panel depicts the distribution as a
function of $\phi$.  In both cases  Eq.~(\ref{eq:MergedSplit})
has been integrated over the entire parameter range of the
correspondingly remaining angle as well as the signal photon energy.
These plots imply that the merged
photons are emitted into a rather small solid angle element: they
predominantly propagate in the $\rm x$-$\rm z$ plane at an outgoing
polar angle of $\theta \approx 2.46{\rm rad}$ (this result is
illustrated in the right panel of Fig.~\ref{fig:FullMergingNumbercon},
where the dashed vector $\vec{k}$ has been chosen correspondingly).
Most notably, for our specific set-up the requirements of momentum and
energy conservation yield a maximum emission of signal photons into
regions where the fields of both the pump and the probe beams
essentially vanish. Such a constellation is most favorable for optimizing the
signal to noise ratio.  Note that a Gaussian beam focussed down to the
diffraction limit, as exemplarily depicted in
Fig. \ref{fig:FullMergingNumbercon} (right panel), has an opening
angle of $1/\pi \approx 0.31{\rm rad}$.  The resulting emission of the
signal into the ``free field'' area together with the comparatively
large number of signal photons makes this scenario an ideal candidate
to experimentally verify the nonlinear nature of the quantum
vacuum. Note that the laser parameters employed here resemble the
parameters of typical petawatt class optical laser facilities which
are {\it currently} in operation. Facilities with even more intense
laser beams are being projected and developed (see, e.g.,
\cite{Vulcan10PW,OmegaEP,ELI,XCELS}).

For completeness let us note that it might prove experimentally
challenging to realize a set-up with two exactly counter-propagating
lasers as considered here, cf. Fig. \ref{fig:FullMergingNumbercon}
(right panel). Even though this set-up provides for a larger number of
merged photons, it might be desirable to avoid counter-propagating
beams for experimental purposes. Keeping the remaining experimental
parameters fixed but taking, e.g., $\theta' = \frac{3\pi}{4}$ and
$\theta''=0.77{\rm rad}$ still yields a total number of
$\mathcal{N}_{\rm Merg} \approx 11.7$ signal photons per laser shot
(for s$'$,f$'' \rightarrow$ s), emitted predominantly at
$\theta=1.92{\rm rad}$. In summary, our merging proposal
  facilitates rather flexible experimental realizations without
  fine-tuning requirements for the geometry of the incoming pump and
  probe beams.

Let us finally compare the all-optical configuration considered here
with the four-wave mixing scenario suggested in
\cite{Lundstrom:2005za,Lundin:2006wu}.  The latter scenario focuses
on the mixing of three incident photon waves from the outset to
address elastic photon-photon scattering with high power lasers.  Both
scenarios have the use of high-intensity lasers in common as well as
the same underlying set of Feynman diagrams. A main difference is that
we consider the merging of two photon waves in a pump field
inhomogeneity which is not restricted to the electromagnetic field of
a propagating laser beam. Our formalism generalizes straightforwardly
to any inhomogeneous pump field.  Moreover, in our approach we can
naturally make contact with the constant crossed-field limit for the
pump field configuration as well as read off the various selection
rules which govern the photon merging process.

%%%%%%%%%%%%%%%%%%%%%%%%%%%
\section{Conclusions}		\label{sec:concl}
%%%%%%%%%%%%%%%%%%%%%%%%%%% 

We have investigated photon merging and splitting processes in
inhomogeneous, slowly varying electromagnetic fields, based on the
three-photon polarization tensor following from the Heisenberg-Euler
effective action. The influence of inhomogeneities appear particularly
promising in the context of high-intensity laser facilities. For the
parameter range of a typical petawatt class laser as pump and a
terawatt class laser as probe, we provide estimates for the numbers of
signal photons attainable in an actual experiment. The combination of
frequency upshifting, polarization dependence and scattering off the
inhomogeneities yields an inherent signal-to-background
separation. This may give rise to successful implementations of
single-photon detection schemes and renders photon merging an ideal
signature for the experimental exploration of nonlinear quantum vacuum
properties.

The central theoretical tool for these results is the explicit
representation of the three-photon polarization tensor at one-loop
order for slowly varying, but otherwise arbitrary electromagnetic
field backgrounds.  This expression allows to analyze in detail the
selection rules for photon splitting and merging in crossed fields. We
have also been able to demonstrate how the well-established
restrictions arising from selection rules in constant background
fields are lifted in inhomogeneous background fields.

The framework laid out in this work is ideally suited to obtain
analytical insights into three-photon interaction processes induced by
vacuum fluctuations in the strong electromagnetic fields generated by
high-intensity lasers. The relevance of inhomogeneities becomes
obvious from the fact that photon splitting and merging processes in
constant background fields are suppressed as $({\cal E}/{\cal E}_{\rm
  cr})^6$ in the ratio of the background field strength $\cal E$ to
the critical field strength ${\cal E}_{\rm cr}$. This consequence of
the Adler theorem can be circumvented by inhomogeneous background
fields which allow for momentum and energy transfers between the probe
photons and the background field. As a result, the suppression is
decreased to only $({\cal E}/{\cal E}_{\rm cr})^2$.

As the our quantitative estimates for the number of attainable signal
photons already yields a decent amount of merged signal photons per
laser shot -- even for already existing state-of-the-art $1{\rm PW}$
class high-intensity laser systems -- we believe that photon merging
can be a good candidate to detect and investigate the optical
nonlinearities of the quantum vacuum for the first time.

%%%%%%%%%%%%%%%%%%%%%%%%%%%
\section*{Acknowledgments}
%%%%%%%%%%%%%%%%%%%%%%%%%%%

The authors would like to thank Matt~Zepf and Malte~C.~Kaluza for
helpful conversations and stimulating discussions. Support by the DFG
under grant No.~SFB-TR18 is gratefully acknowledged.

\end{document}